\documentclass[acmsmall]{acmart}

\usepackage[utf8]{inputenc}
\usepackage{datetime}
\usepackage{adjustbox}
\usepackage{multicol}
\usepackage{mathrsfs}
\usepackage{amssymb}

\usepackage{float}
\usepackage{tabularx}
\usepackage{booktabs}
\usepackage{amsmath,amssymb,amsfonts}
\usepackage{algorithmic}
\usepackage{graphicx}
\usepackage{textcomp}
\usepackage{xcolor}
\usepackage{comment}
\usepackage{array}
\usepackage{multirow}
\usepackage{makecell}
\usepackage{url}
\usepackage{breakurl}
\usepackage{pifont}
\usepackage{balance}
\usepackage[pagewise]{lineno}
\usepackage{hyperref}
\usepackage{subcaption}

\newcommand{\tabitem}{~~\llap{\textbullet}~~}
\newcommand{\cmark}{\ding{51}}%
\newcommand{\xmark}{\ding{55}}%
\newcommand{\revColor}{\textcolor{black}}


\title{Foundations, Properties, and Security Applications of Puzzles: A Survey}

\author{Isra Mohamed Ali, Maurantonio Caprolu, Roberto Di Pietro}
\affiliation{
  \\Division of Information and Computing Technology (ICT) \\ College of Science and Engineering (CSE), Hamad Bin Khalifa University (HBKU) \\ Doha, Qatar \\
  isali@mail.hbku.edu.qa, mcaprolu@mail.hbku.edu.qa, rdipietro@hbku.edu.qa 
}
\date{February 2020}

\acmYear{2020}\acmVolume{0}\acmNumber{0}\acmArticle{0}
\acmMonth{4}

\begin{document}

\begin{abstract}
Cryptographic algorithms have been used not only to create robust ciphertexts but also to generate cryptograms that, contrary to the classic goal of cryptography, are meant to be broken. These cryptograms, generally called puzzles, require the use of a certain amount of resources to be solved, hence introducing a cost that is often regarded as a  time delay---though it could involve other metrics as well, such as bandwidth. 
These powerful features have made puzzles the core of many security protocols, acquiring increasing 
importance in the IT security landscape. The concept of a puzzle has subsequently been extended to other types of schemes that do not use cryptographic functions, such as CAPTCHAs, which are used to discriminate humans from machines. 
Overall, puzzles have experienced a renewed interest with the advent of Bitcoin, which uses a CPU-intensive puzzle as proof of work. \\*
In this paper, we provide a comprehensive study of the most important puzzle construction schemes available in the literature, categorizing them according to several attributes, such as resource type, verification type, and applications. 
We have redefined the term puzzle by collecting and integrating the scattered notions used in different works, to cover all the existing applications. Moreover, we provide an overview of the possible applications, identifying  key requirements and different design approaches. Finally, we highlight the features and limitations of each approach, providing a useful guide for the future development of new puzzle schemes.
\end{abstract}

\maketitle

\section{Introduction}
\label{sec:introduction}
The concept of `\textit{puzzle}' was introduced in the field of security by Merkle in 1978, when he proposed a puzzle to reach key agreement over insecure channels \cite{merkle1978secure}. 
Since the mid-nineties, puzzles have witnessed a growing interest by the research community in a variety of security fields ranging from cryptography and network security to computer performance and bio-metric technologies.  
\par Puzzles constitute the core of many security protocols. They have been proposed as a security tool to achieve various goals, including defending against large-scale attacks, delaying the disclosure of information, creating uncheatable benchmarks, achieving consensus, and differentiating between humans and internet bots.

\par A puzzle is a moderately hard problem that is much easier to verify than to solve. Two key features of puzzles are the asymmetry and adjustability of workload. The solver is forced to dedicate a non-trivial amount of resources to find a solution, while the verifier can check its validity and correctness with a much-reduced effort. 
The adjustability of its hardness enables the verifier to tune the minimal amount of time and resources to be spent by the solver.

\par The process of spending resources to solve a puzzle generally introduces a time delay and an economic cost. The combination of these two effects makes puzzles a powerful tool that can be utilized to limit the capability of an adversary and prevent him from gaining significant influence. The first to observe this phenomenon were Dwork and Naor \cite{dwork1992pricing} in 1992, who proposed a type of puzzles, called \textit{pricing functions}, as a solution to combat spam. 
In a similar context, \textit{client-puzzles} were later proposed by Juels and Brainard \cite{juels1999client} to mitigate denial of service attacks. The general idea of such puzzles is to associate a cost for each resource allocation request by requiring the client to complete a task before the server performs any expensive operation, hence making large-scale attacks infeasible.

 \par One of the areas where puzzles have the most impact is in cryptocurrencies and other emerging technologies, such as blockchain. The idiosyncratic features of the puzzles, known as \textit{proofs-of-work}, have paved the way to the implementation of a fully decentralized peer-to-peer cryptocurrency system. The idea of using puzzles in the creation of a digital-cash payment system has been long investigated \cite{rivest1996payword, dai1998b, szabo2008bit}, but only successfully implemented with the start of 
 the Bitcoin project \cite{nakamoto2008bitcoin} in 2008. The puzzle is used to secure the public ledger of transactions by requiring miners to find a solution before being able to add a block to the ledger. The computational cost imposed by the puzzle prevents a computationally bounded adversary from double-spending transactions or effectively rewriting the ledger. The uniqueness and scarcity provided by the puzzle gives the currency an economic value and enables the process of minting currency \cite{narayanan2017bitcoin}. 

\par Another common adaptation of puzzles is in Bot detection by web-based services. Companies such as Google, Yahoo, and Paypal, use a special type of puzzles, widely known as CAPTCHAs, to verify that the user is not a computer program. 
This type of puzzles enables human identification by using AI-hard problems that, ideally, cannot be solved by machines but can be easily solved by a simple human interaction \cite{von2003captcha}. The human feature provided by this type of puzzles is leveraged to slow down attackers  and to prevent abuse caused by malicious bot programs masquerading as humans.

\par The early puzzle designing approaches concentrated on computational problems that are evaluated by the number of CPU-cycles required to find a solution. An example of such puzzles is the one used in many proposals, including Bitcoin, and was initially introduced by Back \cite{back2002hashcash} in the Hashcash system in 1997. The puzzle requires finding an input to a hash function that produces an output with a specific number of leading zeros. A major drawback of such puzzles is the possible mismatch in the level of processing speeds over time and between different types of processors \cite{dwork2003memory}. This problem was addressed by Abadi et al. \cite{abadi2005moderately} in 2003, who introduced an alternative computational approach that relies on memory-latency, known as \textit{memory-bound functions}. Since memory-latency values are normally more stable than CPU-speeds, most recent systems will solve the puzzle at a similar speed. Another approach was to rely on network latency by Abliz and Tznati \cite{abliz2009guided}. Subsequently, several works that rely on memory and bandwidth were presented in different research areas.

\par Although there have been several proposed construction schemes and a wide-range of applications for puzzles, to the best of our knowledge there has not been any attempt to characteristically distinguish puzzles from other related notions. 
In this paper, we define the term `puzzle' as an umbrella name subsuming all moderately hard functions that are relatively easier to verify than to solve. 
\\

{\bf Contributions} This work can be seen as an extensive introduction to puzzles,  providing the reader with a theoretical background and an overview on the different types of puzzles. Our work aims to fill the gap between the growing works on puzzles and the lack of a comprehensive survey that covers the different types of puzzles. Another aim is to clarify the connections and differences between the terms and notions used to describe a puzzle. We summarize our contributions  as follows: 

\begin{itemize}
    \item We collect and integrate the scattered notions of puzzles into a uniform introductory work. 
    \item We determine the criteria for puzzle categorization. 
    \item We provide an overview of the applications of puzzles and identify the key requirements and challenges faced in each application field.
    \item We examine the different approaches used in the design of puzzles and identify the features and limitations of each one, ideally inspiring the development of novel, effective puzzle schemes.
\end{itemize}

{\bf Roadmap.} The rest of this paper is organized as follows. Section \ref{foundations} provides an introductory overview on puzzles and describes their different types. Section \ref{features} lists the properties and idiosyncratic features of puzzles. Section \ref{applications} provides a survey on the applications of puzzles, states the key requirements for each field, and discusses the viability of puzzles in each application field. Section \ref{schemes} provides an in-depth survey of the state-of-the-art construction schemes and further developments. Finally, Section \ref{conclusion} concludes the survey by summarizing our contributions.

\begin{figure*}
\centering
\includegraphics[width=\linewidth]{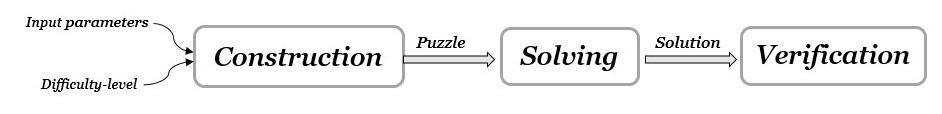}
\caption{Phases of a puzzle scheme.}
\label{fig:phases}
\end{figure*}

\section{Foundations and Background} \label{foundations}
\par Unlike traditional security problems and algorithms, such as cryptograms which ideally cannot be cryptanalyzed, a puzzle is defined as a problem that is meant to be solved \cite{merkle1978secure}. It is easy to verify, such that it is easy to determine if the given inputs produce the given outputs, but moderately hard to solve, such that it is solvable in a reasonable time \cite{dwork1992pricing}. Solving it involves performing a number of operations that require a specific amount of resources, such as CPU cycles, memory, bandwidth, and human's attention. The terms 'easy', 'moderate', and 'hard' are relatively used since their exact definitions depend on the application and implementation of the puzzle. 
\par The solution of a puzzle \textit{in some cases} serves as a proof of work (PoW), in which it demonstrates to one party (the verifier) that the other party (the prover) has performed a specific amount of computational work in a pre-defined time interval \cite{jakobsson1999proofs}. In this sense, the purpose of puzzles differs from the standard cryptographic objective of showing the possession of a secret to proving the ability to expend a certain amount of resources within a certain time interval instead \cite{jakobsson1999proofs}. Anyone without the secret can solve the puzzle but only in one way, which includes dedicating a minimal amount of resources. In the following, we describe the abstract structure of puzzles and present their different categories.

\subsection{Abstract Structure of Puzzles}\label{sub:puzzleStructure}
\par The abstract structure of a puzzle scheme involves two parties, a \textbf{verifier} and a \textbf{prover}. The verifier determines the parameters of a puzzle and checks the correctness and validity of the solution submitted by the prover. The prover solves the puzzle to prove that it is a legitimate party or to obtain a specific reward.
\par A puzzle consists of three main components: \textbf{input parameters}, a \textbf{difficulty-level parameter} and a \textbf{tunable function}. Input parameters are application-related data, such as the message in spam defense or block of transactions in cryptocurrencies. The difficulty-level parameter plays a role similar to that of a security parameter in a cryptosystem \cite{dwork1992pricing}. It determines the hardness of a puzzle and its effectiveness. The issuer should be able to tune the difficulty level flexibly according to the threat level and to accommodate Moore’s Law, such that the tunable function adapts to the increasing amount of computational power and resources over time. 
In general, tuning the difficulty involves adjusting the size of the solution search space or revealing different degrees of the puzzle’s solution. 
\par Any puzzle scheme can be abstracted into three main phases, as illustrated in Figure \ref{fig:phases}: \textbf{Construction}, \textbf{Solving}, and \textbf{Verification}. 
In the Construction phase, a puzzle may be constructed by the verifier or the prover, depending on the scheme type (whether it is interactive or non-interactive). It may also include some offline pre-computations to reduce the online construction cost. This phase provides the two parties the information needed to execute subsequent phases. Once the puzzle is constructed, the prover starts performing the required operations to find a solution and then submits it to the verifier within the specified time interval. Finally, the verification phase involves verifying the validity and correctness of the submitted solution. 

\revColor{The execution of these phases is done through a protocol that can be either interactive or non-interactive. 
The former involves multiple rounds of communication executed by both parties; the prover and the verifier. It terminates by either accepting or rejecting the submitted solution, which is decided by the verifier in the verification phase. 
Non-interactive puzzles involve only one round of communication that is either initiated by the prover, which constructs and solves the puzzle then sends the solution to the verifier, or by the verifier, which constructs a puzzle that does not require an explicit verification and then sends it to the prover. 
The latter is a special type of non-interactive puzzles, known as implicit puzzles, where the verification is determined by the ability of the prover to perform a certain task that can be done in the absence of the verifier, such as in time-lock puzzles \cite{rivest1996time}. We discuss the verification type and the interactivity of puzzles in Section \ref{sub:VerificationType}.}

\subsection{Types of Puzzles}\label{sub:types}
Puzzles can be categorized based on their \textit{application}, the \textit{resources} required to solve them, or the \textit{verification type} of the scheme. The application determines the way in which a puzzle can be utilized. It also defines the requirements and desirable properties of the puzzle scheme. The resources required to solve the puzzle define the metric by which its hardness is measured, whether it is computational steps, memory accesses, memory space or bandwidth \textit{etc}. Finally, the verification type refers to the means by which the solution of a puzzle is used and whether it requires implicit or explicit verification.
\revColor{The rationale behind having more than one categorization is that none of the cited  aspects is  directly related to each other. 
The resource type is not directly determined by the application type, and vice versa. Furthermore, the verification type is determined by the application field and not by the application type itself. Therefore, providing a single categorization that combines any of these aspects may not be possible. In the following, we present the different types of puzzles with respect to the aforementioned aspects and discuss their characteristics as well as the relationships between 
the three aspects: application, resource type, and verification.}

\subsubsection{Application}
\par Puzzles have been relied on by several security protocols in the literature.
Historically, puzzles have been utilized as a \textit{pricing tool} to assign a certain cost for accessing a resource or service, as a \textit{delaying tool} to delay accessing a specific resource, as a \textit{metering tool} to meter the access of a specific resource, as an \textit{identity assignment tool} to achieve consensus in decentralized systems, and as a \textit{human identification tool} to discriminate against humans from bots.
\par In the following, we list the different types of puzzles based on the way they are applied and present a brief description of each. 
\begin{itemize}\small 
    \item \textit{\textbf{Pricing puzzle}}: As defined by Dwork and Naor \cite{dwork1992pricing}, it is a function that is moderately-hard to compute and requires a known lower-bound expenditure of resources. It is not amenable to amortization and cannot be computed more efficiently after some pre-processing. The difficulty of computing a pricing puzzle is leveraged to increase the cost of launching automated large-scale attacks. It is usually applied in contexts where the low cost of using a service leads to abuse.
    
    \item \textit{\textbf{Delaying puzzle}}: Also known as time-lock puzzle \cite{rivest1996time}, it is a moderately-hard function that requires a precise amount of time \textit{(real time, not CPU-time)} to compute. It can only be computed sequentially by performing a deterministic number of computations, and cannot be solved significantly faster with large investments in hardware. The number of computational steps required to find the solution is predetermined by the puzzle issuer allowing him to control precisely when the prover can access the 'locked' resource.
    
    \item \textit{\textbf{Timing puzzle}}: As first introduced by Franklin and Malkhi \cite{franklin1997auditable}, it is a moderately-hard function that requires performing computations incrementally with increasingly large efforts invested. At every stage of the computation, a solver can generate a solution which verifies that a given state is certainly the current state of the computation \cite{boneh2018verifiable}. The number of computation steps is determined in the solving phase instead of being fixed during the construction phase (as in delaying puzzles). The difficulty of solving a timing puzzle is used to ensure the security of a metering/measuring method.
    
    \item \textit{\textbf{AI-hard puzzle}}: As described by Ahn et al. \cite{von2003captcha}, it is a function that can generate and verify problems that a defined portion of the human population can solve but current computer programs cannot solve. Ideally, it is hard for a machine to compute, which allows ensuring that a human is in the communication channel and not a bot.  
\end{itemize}

\par Puzzles provide several features that may be exploited to achieve a specific effect. For example, solving a puzzle requires spending resources hence introducing a time delay that allows the verifier to define when the prover may access the protected service. This feature is exploited by several schemes to achieve the timing effect. Generally, the application of a puzzle is motivated \textbf{\textit{by one or more}} of the following features:
\begin{itemize}\small 
  \item \textbf{\textit{{Computation:}}} This feature requires dedicating a defined amount of resources to solve the puzzle. It is utilized in order to associate a cost to a specific service or activity, such as sending an email or mining cryptocurrencies. 

  \item \textbf{\textit{{Timing:}}} This feature requires the prover to invest a specific amount of time in solving the puzzle. It is utilized to introduce a delay or to measure the time spent on a specific activity, such as visiting a website. 

  \item \textbf{\textit{{Human:}}} As the name implies, this feature requires human interaction to find a solution since it cannot be solved automatically by machines. It is utilized to differentiate between humans and machines. 
\end{itemize}

\par The timing feature is tightly related to the other features, however, schemes which are mainly motivated by time are not affected by the amount of resources used to solve the puzzle. In particular, puzzles that are used as a timing or a delaying function cannot be solved faster using more resources, such as multi-processors in parallel computing. On the other hand, pricing puzzles, which are used as a pricing function, are designed to ensure that the prover performs a certain amount of computational work that introduces a cost and limits the rate of a specific attack to the amount of resources available to adversaries. Unlike timing puzzles, the time required to solve a pricing puzzle is probabilistic, where it has a predictable expected time but a random actual time \cite{back2002hashcash}. Finally, human puzzles, such as CAPTCHAs \cite{von2003captcha}, are the only type of puzzles that has the human feature, in which they are mainly designed to ensure that the prover is a human. We would like to note that human puzzles may be considered as computational puzzles, where the computational work is performed by the human and the dedicated resource is his attention.

\subsubsection{Resource Type}\label{sub:resourseType}
Puzzles are either \textit{CPU}, \textit{memory}, \textit{bandwidth}, \textit{network} or \textit{human} bound (AI hard). 
\begin{itemize}\small
\item \textbf{CPU-bound puzzles:} the computational work is quantified by the number of CPU cycles required to find the solution, which varies vastly in time according to Moore's law, as well as across different machines. 
\item \textbf{Memory-bound puzzles:} the computation is evaluated either by the number of memory accesses or the amount of memory space required to solve a puzzle. Therefore, the computation speed of these schemes is bound by memory latency and bandwidth.
\item \textbf{Bandwidth-bound puzzles:} are evaluated by the amount of bandwidth dedicated to solve the puzzle.
\item \textbf{Network-bound puzzles:} the time required to solve the puzzle is bounded by network latency as the solving process involves sending and receiving packets in a certain order. 
\item \textbf{Human-bound puzzles:} are puzzles that cannot be computed by artificial intelligence but can be easily solved by a simple human interaction. They are evaluated not by the amount of computational resources but by the amount of attention a human dedicates to solve the puzzle.
\end{itemize}

\revColor{In the general case, the resource type is neither directly related to, nor determined by, the selected application type, since different types of resources can be used in the same application field. For example, there exist several implementations and proposals of puzzles in decentralized cryptocurrencies that are bounded by different resources including CPU \cite{back2002hashcash}, memory \cite{biryukov2016argon2}, and human interaction \cite{blocki2016designing}. However, for specific applications, such as Bot detection, the only type of resource that can be utilized  to discriminate between humans and machines is human interaction.
Furthermore, the same type of resource can be used in different applications, such as CPU, which bounds pricing puzzles applied in DoS defense, delaying puzzles applied in time-release cryptography, and timing puzzles applied in uncheatable benchmarks. } 

\subsubsection{Verification Type}\label{sub:VerificationType}
The verification of a puzzle scheme can either be \textit{explicit}, where it is performed by the verifier or \textit{implicit}, where it is determined by the ability of the prover to successfully complete a specific task without the involvement of the puzzle issuer \cite{jakobsson1999proofs}.  \begin{figure*}
\centering
\includegraphics[width=1\linewidth]{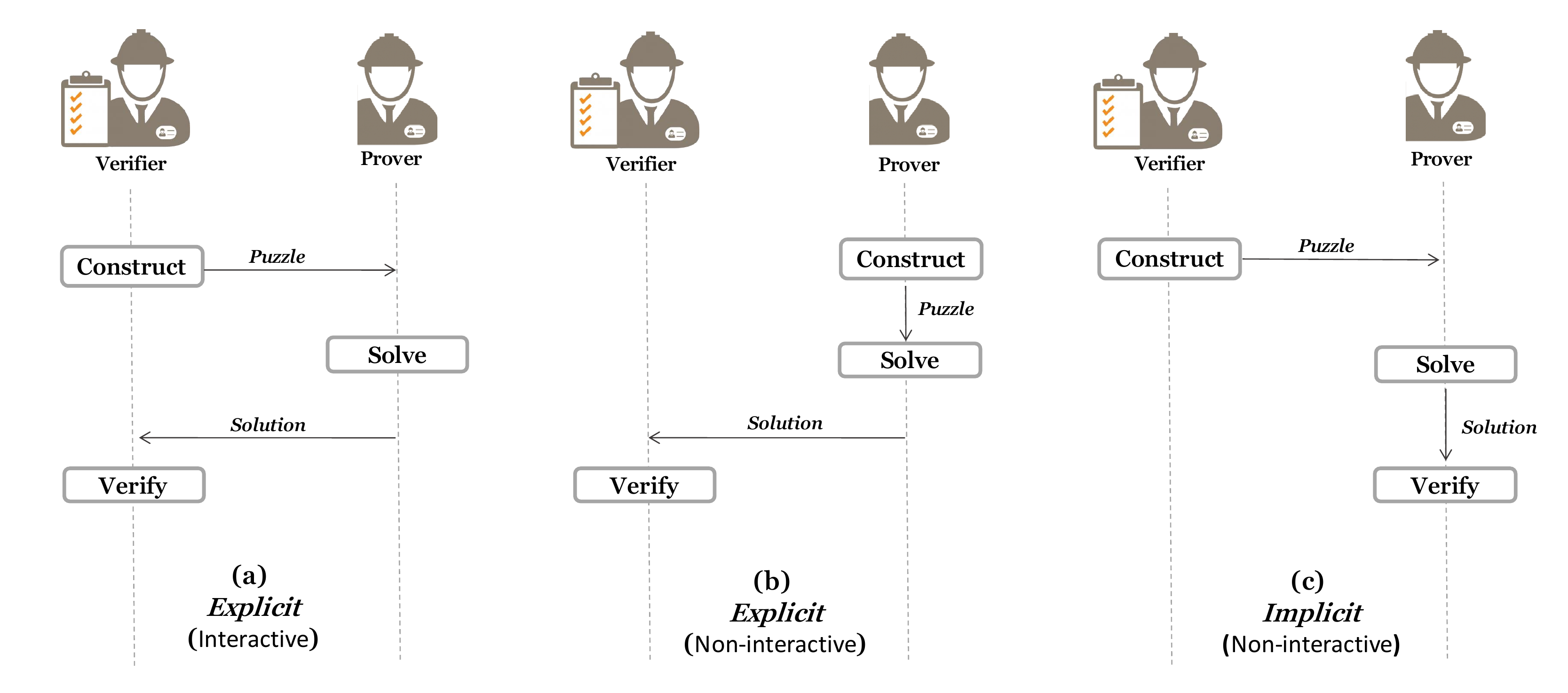}
\caption{Types of puzzle schemes based on the verification and interactivity: (a) is an explicit interactive scheme, (b) is an explicit non-interactive puzzle scheme, and (c) is an implicit non-interactive puzzle scheme.}
\label{fig:verificationTypes}
\end{figure*}

In Figure \ref{fig:verificationTypes}, we illustrate the different types of puzzle schemes based on the verification type and interactivity. \revColor{An explicit puzzle scheme can be executed through both interactive and non-interactive protocols, while an implicit puzzle scheme is only executed through a non-interactive protocol since it requires one round of communication only.} 
\revColor{Consider \textbf{(P, V)} as a two-party protocol through which a puzzle scheme is executed, where \textbf{P} is the prover and \textbf{V} is the verifier. An interactive puzzle scheme is considered as a two-message\footnote{\revColor{We highlight  that such a protocol involves at least two messages, that may increment based on the designed scheme and application field.}} \textbf{(P, V)} protocol, while a non-interactive puzzle scheme is considered as a one-message \textbf{(P, V)} protocol.}

\revColor{In an \textit{explicit} two-message \textbf{(P, V)} protocol, \textbf{V} executes \texttt{\textbf{Construct}} and then sends the generated puzzle \textbf{\texttt{Puzz($I_{n}$,t, k)}} to \textbf{P}, where $I_{n}$ is a set of \textbf{\textit{n}} application related input parameters, \textbf{\textit{t}} is the maximum time required to solve the puzzle, and \textbf{\textit{k}} is the difficulty level parameter. Upon receiving the puzzle, \textbf{P}  executes \texttt{\textbf{Solve}} and then sends the produced solution \textbf{\textit{S}} to \textbf{V}. The protocol terminates by \textbf{V} executing \texttt{\textbf{Verify(S, $I_{n}$,t, k)}} that outputs either \texttt{accept} or \texttt{reject}. }

\revColor{In an \textit{explicit} one-message \textbf{(P, V)} protocol, \textbf{P} executes both \texttt{\textbf{Construct}} and \texttt{\textbf{Solve}}, then sends a message containing both the generated \textbf{\texttt{Puzz($I_{n}$,t, k)}} and the produced solution \textbf{\textit{S}}. Upon receiving the message, \textbf{V} executes \texttt{\textbf{Verify(S, $I_{n}$,t, k)}} that outputs either \texttt{accept} or \texttt{reject}. }

\revColor{Finally, in an \textit{implicit} one-message \textbf{(P, V)} protocol, \textbf{V} executes \texttt{\textbf{Construct}} and then sends the generated puzzle \textbf{\texttt{Puzz($I_{n}$,t, k)}} to \textbf{P}. \textbf{P} then executes both \texttt{\textbf{Solve}} and \texttt{\textbf{Verify(S, $I_{n}$,t, k)}}, where the latter outputs either \texttt{success} or \texttt{failure}.}

\revColor{In the following, we discuss each verification type and provide some examples.}
\begin{itemize}\small 
    \item \textbf{Explicit:} In an explicit puzzle scheme, the solution serves as a way of convincing the verifier that a specific amount of effort is spent. The scheme can either be \textit{interactive} of multiple communication rounds or  \textit{non-interactive} of one communication round, where the prover generates and solves the puzzle in the absence of the verifier. A non-interactive puzzle can be applied in an interactive scheme as a challenge-response protocol, while the converse is not possible. In non-interactive schemes, it is necessary to ensure that the prover cannot effectively control the puzzle generation nor he can precompute the solution. This can be achieved by referencing the puzzle generation to a public source of randomness. \revColor{An example of an explicit puzzle scheme that can be used in both interactive and non-interactive settings is the Hashcash proof-of-work scheme \cite{back2002hashcash}, which is also used in Bitcoin \cite{nakamoto2008bitcoin}}.
    \item \textbf{Implicit:} In an implicit puzzle scheme, anyone, including the prover, can verify a correct solution without the participation of the puzzle issuer \cite{jakobsson1999proofs}. The verification is determined by the prover's ability to carry out a particular task, such as decrypting a message using the key obtained from solving time-lock puzzles \cite{rivest1996time}. An implicit puzzle scheme is considered as a special type of non-interactive puzzle schemes since it requires only one communication round.
\end{itemize}

\revColor{ The type of verification is determined by the specified application field. In particular, explicit verification is required when the solution is used as a proof to \textit{convince} the verifier that the prover did solve the puzzle. This includes pricing puzzles used for DoS defense and cryptocurrencies, AI-hard puzzles used for Bot detection, timing puzzles used to produce uncheatable benchmarks, and delaying puzzles used to produce timestamps. 
Contrarily, the delaying puzzle used in time-release cryptography does not require explicit verification, since the solution reveals a key that is used to decrypt a ciphertext. The solution does not serve as a proof of work, but rather as a way to guarantee that the key is only obtained after a certain amount of computational time has passed. }


\begin{table}[]
\begin{tabular}{@{}l l l l@{}}
\toprule
\textit{\textbf{Categorized by}} & \multicolumn{1}{c}{\textit{\textbf{Application}}} & \multicolumn{1}{c}{\textit{\textbf{Resource Type}}} & \multicolumn{1}{c}{\textit{\textbf{Verification }}} \\ \midrule
\textit{\textbf{Types}} & \begin{tabular}[c]{@{}l@{}}
\tabitem Pricing function\\ 
\tabitem Delaying function\\ 
\tabitem Timing function\\ 
\tabitem AI-hard function\end{tabular} & \begin{tabular}[c]{@{}l@{}}
\tabitem CPU-bound\\ 
\tabitem Memory-bound\\ 
\tabitem Bandwidth-bound\\ 
\tabitem Network-bound\\ 
\tabitem Human-bound\end{tabular} & \begin{tabular}[c]{@{}l@{}}
\tabitem Explicit \\ 
\tabitem Implicit\\         

\end{tabular} \\ \bottomrule
\\
\end{tabular}
\caption{Categorization of puzzles.}
\label{table:types}
\end{table}

\section{Idiosyncratic Features}\label{features}
The fundamental requirement of a puzzle, as first introduced, is that it should be easy to verify but moderately hard to solve. With the evolution of puzzles in various security fields, many properties and desirable requirements were defined to form an effective puzzle. \revColor{In the following, we list these properties and provide a brief description of each one. The first seven properties are the fundamental properties that define a puzzle, which are common among all types of puzzles, while the rest are 
essential for specific application fields, but not for all applications.}

\begin{enumerate}
    \item \textbf{\textit{Asymmetry:}} This property describes the nature of a puzzle, as introduced by Dwork and Naor \cite{dwork1992pricing}. The amount of work required by each party is asymmetric; it should be much easier for the verifier to produce a puzzle and verify a solution than for the prover to find a solution.
    \item \textbf{\textit{Granularity and parameterization \cite{dwork1992pricing}:}} There should be proper parameters that could be adjusted to allow the puzzle to scale with Moore’s law. The verifier should be able to flexibly adjust the puzzle difficulty according to the threat level against the underlying protected service.
    For instance, if the difference between two adjacent difficulty levels is large, then increasing the difficulty would add a huge workload on legitimate parties. Linear-grained puzzles have the highest density of difficulty levels, while exponential-grained puzzles have the lowest density.
    \item \textbf{\textit{Amortization-Freeness \cite{dwork1992pricing}:}} The prover should not be able to produce multiple solutions with a cost equivalent to that of producing one solution.
    \item \textbf{\textit{Independence\cite{dwork1992pricing} /Correlation-free [51]:}} Solving a puzzle or knowing the solution of previous puzzles does not help in solving others.
     \item \textbf{\textit{Efficiency:}} This describes the efficiency of a puzzle measured in terms of cost and overhead introduced to both parties of the puzzle scheme, as follows:
    \begin{enumerate}
       \item \textbf{\textit{Low construction and verification cost:}} It is not enough to maintain a work gap between the verifier and prover, but also to not incur a burden on the verifier that subjects it to resource-exhaustion attacks.
       \item \textbf{\textit{Stateless \cite{juels1999client}:}} A stateless puzzle scheme does not require the verifier to store any information to be able to verify a solution, which is desirable in constrained environments. 
        \item \textbf{\textit{Memory-less \cite{biryukov2017equihash}:}} A memory-less puzzle scheme does not require the verifier to access the main memory for verification. This property is desirable for memory-bound/hard schemes to enable fast verification and allow low-memory verifiers to participate in the scheme.
       \item \textbf{\textit{Minimum interference [51]:}} The operations performed to solve a puzzle should not interfere with concurrently running applications. 
       \item \textbf{\textit{Minimum communication complexity \cite{tang2010non}:}} Refers to the bandwidth and number of rounds required to exchange puzzles and solutions between the two parties. A puzzle scheme should have very low communication complexity and should not add significant traffic to the network. 
    \end{enumerate}
    \item \textbf{\textit{Unforgeability \cite{chen2009security}:}} Initially introduced by Chen \cite{chen2009security}, the prover should be unable to forge a puzzle in a way that allows him to precompute the solution. The lack of this property makes the puzzle ineffective in many applications, such as mitigating DoS attacks.
    \item \textbf{\textit{Freshness \cite{feng2005design}:}} This property is also referred to as tamper-resistance by Feng et al. \cite{feng2005design}, which indicates that the puzzle’s solution is not valid indefinitely and cannot be reused by other provers. In other words, the puzzle scheme should be resilient to replay and precomputation attacks.

    \item \textbf{\textit{Uniqueness \cite{feng2005design}:}}
    Having a unique solution for each puzzle is essential in contexts such as time-release crypto, since the solution is used as a key to decrypt the ciphertext.
    \item \textbf{\textit{Puzzle fairness \cite{abliz2009guided}:}} As defined by Abliz and Znati \cite{abliz2009guided}, the time required to solve a puzzle should be similar for all solvers despite their available resources (CPU, memory and bandwidth). This property eliminates the disparity problem between a powerful attacker and a legitimate prover.
    \item \textbf{\textit{Non-parallelizability \cite{rivest1996time}:}} The puzzle can only be solved in sequential steps and cannot be solved in parallel using multiple machines. This property is necessary for achieving the timing effect and ensuring puzzle fairness. It ensures the effectiveness of a puzzle against high-end adversaries, who utilize parallel computing to solve the puzzle faster.
    \item \textbf{\textit{Deterministic \cite{cai1993towards}(Low-cost variance):}} For delaying and timing puzzle schemes, such as time-lock puzzles \cite{rivest1996time}, the number of operations performed to solve a puzzle should be deterministic in order to control the amount of computing time required. Other types of puzzles may have a probabilistic cost, where solving the puzzle has a predictable expected time but a random actual time. In general, lower variance in cost provides better control over the puzzle difficulty and assures puzzle fairness.
     \item \textbf{\textit{Progress-free \cite{biryukov2017equihash}:}}
     In applications, such as cryptocurrencies, the scheme is required to have a random probability distribution,  
     such that the probability of finding a solution 
     is independent of the amount of effort already spent in solving the puzzle. 
    \item \textbf{\textit{Interactiveness \cite{jakobsson1999proofs}:}} A puzzle can be either interactive or non-interactive. The former requires the verifier to send the challenge (such as client puzzles against DoS), while the latter can be constructed without the need for the verifier's participation (such as PoWs used in blockchain systems). 
    \item \textbf{\textit{Publicly Verifiable \cite{jakobsson1999proofs}:}} In some non-interactive puzzle schemes, such as Hashcash, the verification can be performed by any other party without the participation of the puzzle issuer. This property is essential in decentralized systems, where any participant in the network can verify the solution without requiring a central authority or server.
    
    \item \textbf{\textit{Trapdoor-based \cite{dwork1992pricing}:}} A trapdoor mechanism is used in some puzzles to lower the complexity of puzzle verification by storing secret information that reduces the amount of time needed in solving the puzzle significantly. This property may preclude public auditing, where the verifier has a conflict of interests, such as in web visit metering \cite{franklin1997auditable}. A trapdoor-free puzzle provides trust in decentralized systems, such as cryptocurrencies, by ensuring that the issuer has no advantage in solving the puzzle and cannot forge the proof.
    
\end{enumerate}

\section{Applications}\label{applications}
The idea of using moderately hard problems that are much easier to verify than to solve have been long investigated in various application fields. Puzzles are referred to, in the literature, as proofs-of-work (PoWs), timing functions, delaying functions, cost/pricing functions, AI hard functions and CAPTCHAs. Each term is used to describe the application of a puzzle. Nevertheless, they all share the same property of being moderately hard to solve that a polynomial-time party is capable of finding a solution by dedicating a specific amount of resources. Most puzzles were first designed for a specific application, however, researchers are currently investigating the possibility of designing a multipurpose moderately hard function that can be applied in several security fields \cite{alwen2017moderately}. 
\par In this section, we review the significant applications of puzzles and categorize them based on the way a puzzle is used \textit{(as a pricing, delaying, timing or AI function)}. We also discuss the key requirements and the viability of each type in each application field. 
\revColor{
Given the findings of the study we have conducted on puzzles and their utilization in a wide range of application fields, we conclude that all types of puzzles should be asymmetric, parameterizable, amortization and correlation free, unforgeable, fresh, and efficient \footnote{\revColor{Please refer to Section \ref{features} for the description of each of the listed property/requirement.}}.}
For each application, there are specific requirements and properties that are not as important as in other fields. \revColor{In addition, the level of efficiency in terms of construction cost and the granularity of the puzzle are relative to the application field, where the puzzle's effectiveness in some applications is restricted by the easiness of its construction, by the ability to finely tune its difficulty or both.
In  Table  \ref{table:keyReq}, we present the key requirements for each application field.
\footnote{\revColor{We highlight that Table \ref{table:keyReq} is derived based on the study we have conducted on the application aspect of puzzles, which includes a review and an analysis of the different construction schemes targeting the specified application fields, in addition to, works that study the challenges faced by the employment of puzzles as a security mechanism in some application fields.}}  }

\subsection{Pricing puzzles}
The idea of pricing puzzles is to impose a cost for accessing services that can be easily abused by attackers. The requester of a service is charged with the amount of resources required to find the solution of a problem that is much harder to solve than to verify. In what follows, we present the different security fields that pricing puzzles are applied in and discuss the key requirements and viability of each.
\subsubsection{Key agreement}
The notion of \textit{‘puzzle’} was first introduced by Merkle \cite{merkle1978secure}, in 1978, as a method for key agreement over insecure channels. The objective is to allow any two parties to agree on a secret key that will not be known to eavesdroppers. In this scenario, the protocol initiator plays the role of the verifier, in which he constructs a puzzle consisting of N encrypted keys and verifies the solution submitted by the other party (prover). The other party solves the puzzle by selecting one of the encrypted keys and decrypting it using brute-force, then sending its ID to the verifier. Without knowing which key is mapped to that ID, an eavesdropper must decrypt the N keys at random until he encounters the correct one, which requires an effort of $O(N^2)$. Unlike other puzzle schemes, Merkle’s puzzle is required to be infeasible for any polynomial-time party. This requirement makes the method insecure, however, it offers the feature of workload adjustability. The puzzle issuer can control the solution cost by adjusting the difficulty parameter. This feature is exploited by several works to provide a light-weight pairwise key agreement protocol for resource-constrained environments, such as wireless sensor networks \cite{rasheed2011key,westhoff2011method} and low-energy Bluetooth devices \cite{perrey2011wisec}.

\subsubsection{Spam Defense}
In 1992, Dwork and Naor \cite{dwork1992pricing} suggested using puzzles as an access control mechanism. They introduced the concept of \textit{pricing functions} that increase the cost of sending emails in order to mitigate spam. Their approach is fundamentally an economic one, in which the processing time dedicated to solve the puzzle is a finite resource. Therefore, spammers are limited to the amount of computing resources they can afford, which prevents them from sending emails in bulk. 

\par It is a non-interactive explicit puzzle scheme, where the prover is the sender of an email and the verifier is the recipient. The sender is required to construct the puzzle using time, destination, and message as input parameters, then find a solution and send it along with the email. The recipient verifies the attached solution and accepts the email only if the solution is valid. 

\par In 1997, Back \cite{back2002hashcash} rediscovered the idea and implemented the Hashcash system, which was originally intended for spam and DoS defense, and currently used in Bitcoin \cite{nakamoto2008bitcoin}. Hashcash is a non-interactive trapdoor-free proof-of-work scheme that has an unbounded probabilistic solution cost.

\par Requiring a puzzle for each email would not only disincentivize spammers but also prevent legitimate mass mailing since it requires a significant expenditure of resources. To solve this problem, Dwork and Naor \cite{dwork1992pricing} introduced the idea of a trapdoor-based puzzle, which is much easier to compute given some secret information. Legitimate bulk mail can only be sent cheaply by a centralized trusted authority that holds the secret information. While Hashcash is a decentralized solution that provides better efficiency and resiliency against pre-computation attacks, it does not solve the legitimate mass mailing problem. Nevertheless, it has been deployed in several projects including SpamAssasin \footnote{https://spamassassin.apache.org} 
and Penny Post \footnote{http://pennypost.sourceforge.net/PennyPost}. 

{\bf Key Requirements} To be applied as an anti-spam mechanism, a puzzle scheme should not incur a significant burden on legitimate parties. It should be efficient and non-interactive to avoid requiring the recipient to interact with the sender before receiving an email. It should also consider the sharp disparities across computer systems, as observed by Abadi et al. \cite{abadi2005moderately}, that could make the scheme ineffective against powerful spammers with powerful hardware, restrictively slow for legitimate clients with regular personal computers. They suggest using memory-bound puzzles because memory access speeds differ significantly less than processor speeds. Further developments on memory-bound puzzles in spam defense are proposed in \cite{dwork2003memory, dwork2005pebbling} and discussed in Section \ref{memory-sec}.

\par {\bf Viability} Laurie and Clayton \cite{laurie2004proof} analyzed the viability of pricing puzzles as an anti-spam mechanism and concluded that a universal scheme where every email carries a fixed-cost puzzle is impractical for two reasons. First, a significant proportion of legitimate users are also affected by the added cost preventing them from accessing the service. Second, malicious users can steal CPU cycles by accessing insecure machines, using someone else's resources to solve puzzles. Instead, they suggest incorporating puzzles with other techniques, such as whitelists, to vary the hardness of the puzzle and reduce the burden on legitimate senders. They also suggest using human-bound puzzles, such as CAPTCHAs \cite{von2003captcha}, which are presumably more difficult to steal. On the other hand, several research papers have demonstrated the feasibility of pricing puzzles against spam when used in parallel with other techniques, such as reputation systems \cite{liu2006proof}.

\subsubsection{DoS Defense} Client puzzles are a type of pricing puzzles used to defend against denial of service (DoS) attacks. Namely, resource depletion attacks that prevent the victim from processing legitimate requests for a service in a server-client setting. Before allocating any resources for a given request, the server requires the client to commit a portion of its resources by solving a puzzle. Requests that do not include a correct solution are dropped. Legitimate clients may experience a degradation in service, however, attackers are unable to send a large number of requests simultaneously due to the time delay introduced by the puzzle. 
\par The two key features of a puzzle that qualify it as an anti-DoS mechanism is workload adjustability and asymmetry. The former allows the server to tune the difficulty level of the puzzle according to the current threat level, by initially setting it to zero when there are no attacks and increasing it as the intensity of the attacks increases. While the latter shifts the workload from the server to the client as, it is much easier for the server (verifier) to construct and verify a puzzle than for the client (prover) to solve the puzzle.

\par The first construction scheme of client puzzles was proposed by Jules and Brainard \cite{juels1999client} as a countermeasure to connection depletion attacks. Aura et al. \cite{aura2000resistant} later improved and generalized the design of the puzzle to employ it in any authentication protocol. Many construction schemes have been subsequently proposed to protect various types of services including network IP and TCP channels \cite{feng2005design, mcnevin2004ptcp, wang2003defending, wang2004mitigating, noureddine2018revisiting}, TLS \cite{dean2001using, nygren2016tls}, capability-granting channels \cite{parno2007portcullis}, and key agreement protocols \cite{stebila2009towards}. 
 Despite the various techniques used by these construction schemes, they all must satisfy the fundamental properties described by Feng et al. \cite{feng2005design}. 

{\bf Key Requirements} 
For a puzzle to be applied in DoS defense, it should be efficient enough to guarantee the availability of the puzzle distribution service and avoid subjecting the scheme itself to a DoS attack. It should be resilient to precomputation attacks where the puzzle solution indicates that the computational effort was recently spent. Furthermore, robust authentication mechanisms must be employed to prevent attackers from spoofing puzzles and disabling the server by falsely triggering the puzzle mechanism against it \cite{feng2003case}. 
\par The granularity of the puzzle, which represents the density of difficulty-levels, should be high in order to allow the server to finely control the amount of computational effort spent by the client. Finally, forcing all clients to solve puzzles before allowing access is crucial to mitigate the attack, however, not all clients have the same capabilities. Therefore, the scheme must also consider low-power clients and adjust the level of puzzle complexity to match the client's capabilities in order to achieve puzzle fairness \cite{feng2003case}.

{\bf Viability} The main challenge faced in deploying puzzles as a DoS countermeasure is determining and setting the appropriate difficulty level that limits the abilities of attackers but not legitimate parties. From an economic perspective, the effectiveness of a pricing puzzle can only be achieved when the amount of work required from legitimate clients and attackers differ significantly \cite{laurie2004proof}. The construction scheme must identify and discriminate against known malicious behavior \cite{feng2003case}. Most client puzzle schemes set the difficulty based on a single metric, such as the load on the system \cite{juels1999client, dean2001using, parno2007portcullis}, the rate at which the client sends requests \cite{feng2005design, kaiser2008mod}, or the level of demand for the service \cite{wang2003defending, wang2004mitigating}. However, using a single metric has been proven to be insufficient \cite{gligor2003guaranteeing}, as it provides clients weak access guarantees at high per-request overhead. 

\par The viability of client puzzles remains an open question. Issues such as constructing puzzles efficiently while ensuring non-parallelizability, adjusting the puzzle difficulty to prevent subversion while maximizing  server utilization, and attaining equitable fairness must all be addressed to incentivize its deployment.

\subsubsection{P2P Systems and Cryptocurrencies}\label{subsub-p2p} In the context of decentralized systems, pricing puzzles are used to address various security issues including Sybil \cite{douceur2002sybil} and collusion \cite{reiter2009making} attacks, achieving consensus in a Byzantine setting, and incentivizing correct behaviour by requiring participants to submit a puzzle solution that serves as a \textit{proof-of-work} and then rewarding them for participation. 
This type of puzzles is utilized by important applications in creating decentralized cryptocurrencies, such as the recent systems Bitcoin \cite{nakamoto2008bitcoin}, Ethereum \cite{wood2014ethereum}, and Litecoin~\cite{lee2011litecoin}, or prior schemes such as the Micromint system of Rivest and Shamir \cite{rivest1996payword}. 

Decentralized dynamic systems are highly vulnerable to the Sybil attack \cite{douceur2002sybil}, whereby the attacker exploits the low cost of forging multiple identities that allows him to control a substantial fraction of the system and execute further attacks to subvert the system. Pricing puzzles have been long investigated as a decentralized Sybil defense mechanism in a variety of p2p network settings and overlays, including structured \cite{aspnes2005exposing, borisov2006computational, rowaihy2007limiting, li2012sybilcontrol}, and unstructured \cite{ nakamoto2008bitcoin, wood2014ethereum} overlays. The idea is to impose a computational cost on maintaining an identity within the system, hence limiting the proportion of Sybil nodes to the proportion of resources that an adversary can control per time unit. 

\par In p2p identity systems, such as SybilControl \cite{li2012sybilcontrol}, the puzzle is used as a distributed admission control mechanism that grants nodes the permission to join and stay functional in the system. In this protocol, all nodes are required to periodically solve a unique puzzle and collectively verify solutions of other nodes. If a node fails to compute the puzzle within the specified time interval, its identity gets revoked. 

\par Decentralized cryptocurrencies, such as Bitcoin \cite{nakamoto2008bitcoin} and Ethereum \cite{wood2014ethereum}, utilize pricing puzzles to achieve several goals at once. They exploit the scarcity and uniqueness provided by pricing puzzles to create economic value and mint crypto-currency. More importantly, they use the pricing puzzle as a key component in the blockchain protocol to achieve consensus and prevent double spending \cite{narayanan2017bitcoin}.

\par Blockchain is described as a cryptographic data-structure in which a transaction ledger, that is shared and agreed on by all nodes of the network, is recorded. Compared with the original design of identity pricing puzzle schemes \cite{aspnes2005exposing, borisov2006computational}, the puzzle in a blockchain network is not used in the identity verification of participating peers. Instead, the peers are expected to collectively verify puzzle solutions broadcasted by other peers in order to determine who's block will be considered as the next block in the chain. The Sybil and double-spending attacks are mitigated by associating a computational cost to the process of adding a block to the chain. A comprehensive survey of the consensus protocol design and the effects of blockchain networks is presented by Wang et al. \cite{Wang}. 

\par The cryptocurrency is created by nodes, known as `\textit{miners}', through a process that involves processing a block of transactions and finding a solution to a pricing puzzle that makes the block valid. Once a miner solves the puzzle, it broadcasts the proposed block to all other nodes, and receives a reward only if the block is accepted by the majority of nodes. The validation of that block is done independently by each node in the network, which includes verifying the puzzle solution and correctness of each transaction within the block. Miners are incentivized to act honestly since incorrect puzzle solutions would result in the rejection of the block by the majority of nodes, hence losing the reward and wasting the effort spent in solving the puzzle. Further technical details on Bitcoin and digital currencies is presented in \cite{tschorsch2016bitcoin}.

\par The rate at which blocks are appended to the ledger is determined by the difficulty of the puzzle. It is adjusted such that, it is hard enough for an adversary to interfere and alter the system, but easy enough for miners to construct new blocks and unify their views of the public ledger. The robustness of the consensus protocol relies on the assumption that more than 51\% of the computational resources are possessed by honest participants who follow the longest chain rule. Formal analysis of blockchain's security under different network assumptions appear in \cite{miller2014anonymous, garay2015bitcoin, pass2017analysis, garay2017proofs}.

{\bf Key Requirements} Efficiency is one of the key requirements of a puzzle scheme to be applied in dynamic decentralized P2P systems. The puzzle should be easy to construct and verify, stateless, and compact in order to provide scalability for the underlying protocol. Given the lack of a trusted third party in such environments, the puzzle should be \textit{non-interactive}, \textit{trapdoor-free}, and \textit{publicly verifiable}. Any node in the network should be able to efficiently verify the puzzle solution of any other node without the access to a trapdoor or any secret information. The trapdoor-free property is essential to provide trust in the system. 
\par The freshness property of the puzzle should be ensured at the execution phase \cite{Wang}. In particular, the puzzle solution should be non-reusable and unpredictable such that, the computational work is guaranteed and the proof is unforgeable. Furthermore, fairness should be ensured, such that the probability of finding a solution is directly proportional to the computational power of the node at any given time. This is crucial for cryptocurrencies since solving the puzzle has an economical value. Finally, the hardness of the puzzle should be adjustable in order to adapt to the changing scale and settings of the p2p network.

{\bf Viability}
Notably, the most impactful application of pricing puzzles to this date is in the implementation of permissionless p2p systems, namely, blockchain emerging technologies. Although examples such as Bitcoin and Ethereum demonstrate the success of pricing puzzles in practice, several concerns are raised by the research community regarding the stability of these systems \cite{bonneau2015sok}, the high power consumption \cite{luu2015power} and wastage of resources which increases proportionally to the system's popularity \cite{BitcoinWaste}. Furthermore, the parallelizability nature of the utilized puzzle schemes allows nodes to increase their voting power by using customized hardware (such as ASICs) that solves the puzzle substantially faster. This development in hardware subverts the pricing puzzle approach and implies several threats. In particular, it diminishes the democratic value of decentralized cryptocurrencies by suppressing low-end nodes \cite{tschorsch2016bitcoin}, and enable powerful nodes to collude and alter the system \cite{eyal2018majority}. Several puzzle schemes have been proposed to address these issues including, ASIC-resistant puzzles \cite{biryukov2017equihash, ren2017bandwidth}, non-outsourceable puzzles \cite{miller2015nonoutsourceable}, useful puzzles \cite{ball2018proofs, ball2018proofs, miller2014permacoin}, and eco-friendly puzzles \cite{dziembowski2015proofs, blocki2016designing}. 


\subsection{Delaying and Timing puzzles}
\par The timing feature of a puzzle is exploited by several schemes either to slow down attackers, to lock resources for a precise amount of time, or to measure the time spent accessing a resource. Unlike pricing schemes, delaying and timing schemes are concerned with making \textit{CPU time} and \textit{real time} agree with an approximate precision. They achieve this by using inherently sequential problems that have a deterministic solution cost. It is important to note that the solution time of these puzzles is \textit{approximately controllable} since different computer systems operate at different speeds. In what follows, we discuss the different application fields of these types of puzzles. 

\subsubsection{Uncheatable Benchmarks and Auditable Metering}
\par The first proposal in designing a puzzle that requires solving an inherently sequential problem seems to appear in the application of hardware benchmarking in 1993 by Cai et al. \cite{cai1993towards, cai1998making}. 
The idea is to validate the performance of specific hardware by having it compute a puzzle that reflects its computation power. The asymmetry feature provided by puzzles allows customers with low-end machines to verify the benchmark of high-end computer vendors. A customer provides the hardware vendor with a puzzle, who finds a solution and submits it as a verification of the claimed hardware performance. The customer then verifies the puzzle and checks that the solution time is indeed within the claimed bound. The soundness of the puzzle scheme would guarantee that the vendor could not cheat by optimizing his code or modifying it.

Franklin and Malkhi \cite{franklin1997auditable} proposed the idea of metering client accesses via a timing puzzle. They presented a lightweight solution to the problem of forged client website visits. The timing puzzle requires an incremental amount of computations which makes forging a large number of client visits expensive and time-consuming. The solution of the puzzle serves as evidence that a specific amount of time has passed. At each webpage visit, the client is asked to compute a timing function that requires performing repeated hashing incrementally until the end of the visit. The result is then sent to an auditing proxy which verifies its correctness. 
The main drawback of their construction scheme is the requirement of reconstructing the puzzle for accurate verification, which was later addressed by Chen and Mao \cite{chen2001auditable}.

{\bf Key Requirements} In addition to being non-parallelizable and deterministic, a timing puzzle must provide the ability of applying it incrementally to reflect the actual continuous-time being spent. The solver should be able to produce extendable solutions, where the puzzle difficulty is not set in advance but is incremental with the time spent in computing it. Forging access duration or performance records should require a known amount of resources that increases proportionally to the amount of forgery. For public auditing, the puzzle should be trapdoor free, in which there is no shortcut available for the prover to find a solution with fewer resources.

{\bf Viability} Timing puzzles may be considered as an unreliable metering method due to the existence of several uncontrollable factors that may affect the accuracy of real-time measurement, including network delay, bandwidth, and computational power \cite{blundo2004software}. Moreover, the existing timing puzzle schemes only offer lightweight security requiring precise limits on the adversary's processing speed. 

\subsubsection{Time-Release Cryptography and Time-stamps}
\par The idea of time-release cryptography is to "send data into the future" by encrypting a message that can only be decrypted after a pre-defined amount of time has passed. In 1996, Rivest et al. \cite{rivest1996time} implemented this idea using time-lock puzzles, which can be applied to delay sealed-bid auctions, digital cash payments, and key escrow. They have also been proposed for other applications such as, enabling offline submission \cite{jerschow2010offline}, providing pseudonymous secure computation \cite{katz2014pseudonymous}, constructing a two-round concurrent non-malleable commitment \cite{lin2017two}, and supporting digital forgetting \cite{amjad2018forgetting}.

\par Time-lock puzzle is a cryptographic primitive that allows locking data, making it accessible only after a certain delay. It is an implicit trapdoor-based puzzle scheme, where the solution reveals the key that decrypts the encrypted data. In \cite{rivest1996time}, the puzzle issuer selects the desired time delay and constructs a modular exponentiation puzzle that can only be solved by performing \textit{t} modular squaring operations sequentially. The number of squaring operations can be exactly controlled hence providing the ability to finely tune the difficulty of the puzzle. The verification is not explicitly required, however, can be done more efficiently by anyone who can access the trapdoor which is created in the construction phase.  

Inspired by time-lock puzzles, Mahmoudy et al.\cite{mahmoody2013publicly} introduced the concept of \textit{proof of sequential work} (PoSW), an explicit time-lock puzzle scheme which enables the solver to prove that a specific number of computation steps was performed sequentially on a given challenge. The solution of the puzzle is publicly verifiable and indicates that an approximate number of time units have passed since receiving the puzzle. They propose using PoSW to produce \textit{relative} timestamps for documents, whereby a timestamp \textit{d} of a document at time \textit{T} proves the existence of that document at time $T-d$. A stamper specifies the desirable duration \textit{d}, constructs and solves the puzzle by using the document as an input parameter. The verifier then checks the validity of the timestamp by verifying the solution of the puzzle.

\par Further developments on such construction schemes have been proposed to provide better efficiency \cite{cohen2018simple} and uniqueness \cite{pietrzak2018simple}. Uniqueness is important to guarantee that the solver cannot produce multiple solutions at the same cost of one. Boneh et al. \cite{boneh2018verifiable} study the problem of constructing delaying puzzles, which they refer to as \textit{'verifiable delaying functions'}, and present further applications of such puzzles, including randomness beacons, proof of replication and resource-efficient blockchains.

{\bf Key Requirements}  The key requirements for such applications are \textit{non-parallelizability} and hardware-independence. The solution time should not depend on the amount of hardware being used. A solver who uses a large amount of investments in hardware, namely parallel computing, should not be able to find a solution substantially faster than the pre-determined time. Finally, the time required to construct and verify the puzzle should be much less than the solution time. 

{\bf Viability} The main challenge that may hinder the deployment of a delaying puzzle in a specific application is the requirement of exact guarantees on the precision timing, which cannot be achieved due to the existence of variations in the speed of single computers. For other settings where there are no trusted third parties available, the puzzle construction schemes proposed in \cite{rivest1996time} are unsuitable since generating the puzzle requires knowing the (\textit{secret}) puzzle solution in advance and verification requires accessing a trapdoor. Although the schemes proposed in \cite{mahmoody2013publicly, cohen2018simple} provide public verifiability, their security is only proved in the random oracle model and the uniqueness of the produced solution is not guaranteed.
\subsection{AI-hard puzzles}
AI-hard puzzles, widely known as CAPTCHAs, are intended to ensure the presence of a human in a communication channel by using hard AI problems that differentiate between humans and bots. They are puzzles that are easy-to-solve for humans but hard-to-solve for automated computer programs. In the following, we present the main security fields in which human-bound puzzles are used and discuss their key requirements. 

\subsubsection{Bot Detection in Web-based Services}
The first to suggest using a reverse Turing test for verifying that a human is the one requesting a service over the web was Naor \cite{naor1996verification} in 1996. Several practical examples were then proposed and developed \cite{lillibridge2001method, baird2003pessimalprint, ramanathan2013spam}. In 2000, Von Ahn et al. \cite{blum2000captcha} introduced the notion of a CAPTCHA and provided a formal framework that models it as a hard AI problem in \cite{von2003captcha}. CAPTCHA can be considered as a trapdoor-based explicit puzzle scheme whose difficulty is based on an AI problem that can only be solved by a human. A typical CAPTCHA puzzle is constructed by first generating a random target solution, such as a text or an image, and then performing distortion techniques to that solution in order to make it hard for computers to solve the puzzle. The target solution, which is generated in the construction phase, is later used by the verifier to check the correctness of the solution submitted by the prover.
\par CAPTCHAs are used by Google and many other popular web-based services to mitigate abuse caused by malicious bot programs masquerading as humans. Such abuse includes automated account registration, password guessing attacks, systematical database mining, and massive voting in polls. Before being able to access the service, a client is required to prove that he is a human by solving a given instance of the puzzle. The human feature provided by this type of puzzles slows attackers down and prevents them from sending a large number of fake requests generated by automated programs. 
\par Many variations of construction schemes exist in the literature which are based on different AI problems, including reading distorted text \cite{von2008recaptcha}, audio recognition \cite{lazar2012soundsright, meutzner2015constructing}, human face recognition \cite{goswami2014facedcaptcha}, and emergent-image recognition \cite{gao2015emerging}. A comprehensive review of the different categories and sub-applications of CAPTCHA is presented by Hidalgo and Alvarez \cite{hidalgo2011captchas}. Despite the many variations of AI-hard puzzles, such schemes are based on the hypothesis that the underlying AI problem cannot be solved by the adversary's machine more accurately than what is currently known in the AI field \cite{von2003captcha}. 

{\bf Key Requirements} To provide both practicality and usability, the puzzle should be easy-to-construct for the puzzle generator machine and easy-to-solve for the human solver. Humans should be able to solve the puzzle effortlessly with a negligible error rate. The hardness of the puzzle should be adjustable such that both robustness and usability are guaranteed even with the advance of technology. The best known programs for solving the underlying AI hard problem should fail on a non-negligible portion of the puzzles, despite that the method of constructing the puzzle instances is known \cite{naor1996verification}. Finally, the puzzle scheme should be fair that does not discriminate against disabled people and involve different sensory abilities, including hearing and vision.

{\bf Viability} The main issue that faces the deployment of AI-hard puzzles in web-based services is setting the appropriate difficulty that ensures both resistance against machine-learning attacks and human usability. The same distortion methods used to make the puzzle unsolvable by machines can also significantly degrade human usability \cite{yan2008usability}. CAPTCHAs are often hard for humans to solve due to a number of demographic factors such as age, language, and education \cite{bursztein2010good}. Furthermore, this type of puzzles fails to recognize people with visual and hearing impairments as humans, which prevents them from accessing the underlying protected web service \cite{hidalgo2011captchas}. 
These usability issues may drive customers to abandon services that deploy CAPTCHAs resulting in financial losses for those companies. 

\par Although many construction schemes have been successfully attacked using machine-learning techniques \cite{yan2008low,bursztein14end,mori2003recognizing, xu2012security}, a significant gap between human intelligence and the current artificial intelligence still exists. However, this does not ensure the security and effectiveness of the puzzle scheme as it is still vulnerable to human relay attacks, whereby the puzzle is outsourced to paid human-solvers \cite{danchev2008inside}.
Motoyama et al.\cite{motoyama2010re} analyzed the behavior and dynamics of CAPTCHAs from an economic perspective. They conclude that CAPTCHA is a low-impact mechanism that reduces the attacker’s profitability hence minimizing the cost and legitimate user impact of more expensive secondary defenses. However, they may be ineffective in scenarios where the profit gained from launching the attack is much greater than the cost associated with paying humans to solve the puzzle.

\subsubsection{Decentralized Cryptocurrencies}
Recently, Blocki and Zhou \cite{blocki2016designing} introduced the concept of \textit{proof-of-human-work} (PoH), a human-in-the-loop puzzle that is publicly verifiable and can be used to build a decentralized cryptocurrency system. It is a non-interactive explicit puzzle scheme that can only be solved with sufficient human assistance. Unlike the traditional standalone CAPTCHA, the solution of the puzzle is unknown to the puzzle-generator machine and the difficulty is adjustable.
\par The scheme involves two types of puzzles, a CAPTCHA and a pricing puzzle similar to that used in Bitcoin. To produce a valid block, the miner is required to first prove that he is a human by solving a CAPTCHA instance, then use the obtained solution as an input parameter to the pricing puzzle. The CAPTCHA instance is constructed obliviously using the \textit{universal samplers} developed by Hofheinz et al.~\cite{hofheinz2016generate}. 
The instance is generated along with a verification tag, and the corresponding solution remains concealed in the obfuscated program.


This feature provides public verifiability, where anyone in the network can verify the submitted solution. The difficulty of the puzzle is adjusted by having the verifier reject a valid solution with a certain probability so that the human miner 
have to generate and solve a specific number of puzzles to produce a valid proof-of-human-work. 
\par The human feature provided by the AI-hard puzzle is exploited to satisfy properties that traditional cryptocurrencies such as Bitcoin lack, which are eco-friendliness, usefulness, and centralization-resistance. Eco-friendliness and usefulness are achieved by relying on human effort instead of computational power. The human effort may involve performing educational \cite{khot2009icaptcha} or productive \cite{hwang2012spelling} tasks in order to avoid wasting human cycles. While centralization-resistance is achieved by ensuring fairness, where any two humans are capable of performing a similar amount of work to solve the puzzle. 

{\bf Key Requirements}
In the context of cryptocurrencies, the AI-hard puzzle should be efficient and trapdoor-free, such that it is easy for a machine to construct, but difficult for any machine (\textit{including the puzzle-generator machine}) to solve without sufficient human assistance. It must be non-interactive and publicly verifiable such that a machine can easily verify the puzzle solution and ensure that the issuer does not already have the solution without any human assistance.
Finally, the puzzle scheme should provide a sufficient density of difficulty-levels to allow adjusting its hardness according to the changing scale and settings of the system.  

{\bf Viability} The viability of AI-hard puzzles in decentralized cryptocurrencies is affected by two main challenges. First, a trapdoor-free construction scheme whereby the solution is unknown to any party throughout the generation of the puzzle is required to provide trust in the system. This challenge is addressed by the authors of HumanCoin system \cite{blocki2016designing} using indistinguishably obfuscation (IO), however, the current development achievements in IO do not provide a practical solution hence their approach is currently impractical. Furthermore, their proposed system requires an initial trusted setup phase since the party generating the system may embed a trapdoor allowing it to produce coins without involving any human work. Second, the security and stability of the system rely on the hardness of the underlying AI problem. The life of such cryptocurrency is anticipated to be shorter than other cryptocurrencies that depend on cryptographic primitives since AI breakthroughs are achieved more frequently. Therefore, achieving and maintaining trust using AI-hard puzzles in such cryptocurrency systems remains an open question.

\begin{table}[]
\resizebox{\textwidth}{!}{\begin{tabular}{|l|c|c|c|c|c|c|c|c|c|c|}
\cline{2-10}
\multicolumn{1}{c|}{\textit{\textbf{}}} & \multicolumn{9}{c|}{\textit{\textbf{Key Requirements}}} \\ \hline
\multicolumn{1}{|l|}{\textit{\textbf{Application Field}}}  & 
\textit{\textbf{\begin{tabular}[c]{@{}c@{}}{Easy to}\\{construct}\end{tabular}}} & 
 \textit{\textbf{\begin{tabular}[c]{@{}c@{}}{Fine-}\\{grained}\end{tabular}}} & 
 \textit{\textbf{\begin{tabular}[c]{@{}c@{}}{Deterministic}\\ {computation}\end{tabular}}} & 
 \textit{\textbf{\begin{tabular}[c]{@{}c@{}}{Non-}\\ {parallelizable}\end{tabular}}} & 
 \textit{\textbf{\begin{tabular}[c]{@{}c@{}}{Non-}\\ {interactive}\end{tabular}}} & 
 \textit{\textbf{\begin{tabular}[c]{@{}c@{}}{Publicly}\\{verifiable}\end{tabular}}} & 
 \textit{\textbf{\begin{tabular}[c]{@{}c@{}}{State}\\{-less}\end{tabular}}} & 
 \textit{\textbf{\begin{tabular}[c]{@{}c@{}}{Trapdoor}\\{-less}\end{tabular}}} & 
 \textit{\textbf{\begin{tabular}[c]{@{}c@{}}{Fair}\\{}\end{tabular}}} \\ \hline 
 Spam defense & \checkmark &  &  &  & \checkmark &  &  &  &  
 \\ \hline DoS defense & \checkmark & \checkmark &  &  &  &  & \checkmark &  & \checkmark  
 \\ \hline Cryptocurrencies & \checkmark &  &  &  & \checkmark & \checkmark & \checkmark & \checkmark &
 \\ \hline Delayed Disclosure &  & \checkmark & \checkmark & \checkmark &  &  &  &  & 
 \\ \hline Auditable metering & \checkmark &  &  &  &  & \checkmark &  & \checkmark & 
 \\ \hline Uncheatable Benchmarks &  &  & \checkmark & \checkmark &  &  &  & \checkmark &  
 \\ \hline Bot Detection & \checkmark &  &  &  &  &  &  & & \checkmark \\ \hline
\end{tabular}}
\\
\centering
\caption{Key requirements of puzzles for each application field.}
\label{table:keyReq}
\end{table}

\section{Construction Schemes} \label{schemes}
As discussed previously, puzzles may be categorized based on several aspects including the ways they are applied, the type of verification, and the type of resource that bounds the scheme. In this section, we categorize them based on the resource that bounds the puzzle scheme which includes: CPU, memory, and bandwidth. We survey state-of-the-art puzzles by focusing on the type of construction scheme and present further developments and improvements on these schemes. 
\subsection{\textbf{CPU-Bound Schemes}}\label{sub-cpu-bound}
\subsubsection{Merkle Puzzles}
\par Computational puzzles were first introduced by Merkle \cite{merkle1978secure} as a method to establish a secure key agreement over insecure channels. The main designing goal was to create a work gap between legitimate parties and passive adversaries in order to guarantee secure key distribution. The puzzle is based on symmetric cryptography and is constructed by generating N encrypted keys, each having a unique ID. The other legitimate party solves the puzzle by selecting one of the keys and decrypting it by brute-force then submitting its ID. To know the key, an eavesdropper must decrypt all N keys since he cannot determine which of the IDs is mapped to the key. The puzzle’s difficulty is adjusted by adjusting the size of the key used to perform the encryption. As a key agreement protocol, this scheme is impractical and insecure, since the optimum work gap that can be achieved using this method, as proven by Barak and Mahmoudy \cite{barak2009merkle}, is quadratic. However, it was the building block towards public key cryptography. Merkle puzzles are different than the following puzzles as both parties are required to perform a similar amount of work O(N), hence it is not asymmetric. 
\subsubsection{Pricing functions}
\par Dwork and Naor \cite{dwork1992pricing} were the first to suggest using puzzles as an access control mechanism to combat email spam. They used the puzzle as a ‘pricing function’ to assign a computational cost to resource allocation requests. Their goal is to increase the cost of sending an email for spammers by forcing them to compute a unique puzzle for each recipient. They proposed three puzzle schemes. The first scheme is based on factoring a square root and the other two are based on digital signature schemes with smaller security parameters. The first is verified by squaring the submitted solution, while the digital signature based schemes are verified using trapdoors to reduce the verification cost. The availability of trapdoors in their scheme is essential, not only to reduce the verification cost but to allow legitimate mass mailing by specific authorities. Solving the puzzle requires forging a signature without actually breaking a private key using a moderately hard algorithm such as, the Pollard algorithm. The proposed schemes satisfy the amortization-freeness property, but are prone to pre-computation attacks which weakens the effectiveness of the puzzle. Furthermore, similar to previous schemes, it suffers from inefficiency requiring relatively high construction cost.

\subsubsection{Time-lock and Delaying functions}
Rivest et al.\cite{rivest1996time} rediscovered the idea of puzzles to implement time-release cryptography and introduced the notion of \textit{time-lock puzzles}, which are computational puzzles that can only be solved in a precise amount of time. Their designing goal was to create a puzzle that can only be solved sequentially by performing a deterministic number of operations. The puzzle is used to allow encrypting a message that can only be decrypted (unlocked) after a predetermined period of time specified by the puzzle issuer elapses. This type of puzzles is different than traditional PoW schemes, in which the solution of the puzzle serves as a decryption key rather than a method of convincing the verifier. In particular, the verification is not performed by the verifier but determined by the ability of the prover to decrypt the encrypted message using the solution of the puzzle.

\par The hardness of Time-lock puzzles relies upon the infeasibility of factoring large integers. To construct the puzzle, the issuer first sets its parameters by generating a composite modulus N as the result of multiplying two large private prime numbers \textit{p} and \textit{q}. It determines the time required to solve the puzzle \textit{$t = TS$}, where \textit{S} is the number of squaring operations current machines can run per second, and \textit{T} is the desired time specified by the puzzle creator. The issuer then generates the puzzle by encrypting a message \textit{m} with key \textit{K} and encrypting the key as $C_{k} = K + {a^2}^t mod\ N$. The puzzle is given as (\textit{N, a, t, $C_{k}$, $C_{m}$}) and the solution of the puzzle reveals the key that decrypts the message. Solving it requires performing \textit{t} modular squaring sequentially starting with \textit{a}. The difficulty of the puzzle is adjusted by increasing or decreasing \textit{t}, which provides fine granularity. Knowing $\Phi (N)$, the verifier can use the trapdoor given by Euler’s function to compute the solution in $O(\log N)$ modular multiplications. 
\par Time-lock puzzles guarantee non-parallelizability and deterministic computation, which provides fine-grained difficulty adjustment. However, the requirement of generating large prime numbers and performing modular exponentiations for every puzzle is resource-exhausting, which makes it impractical and unsuitable for resource-constrained environments. 
\par {\bf Efficiency.} To reduce the construction and verification cost, Karame and Capkun \cite{karame2010low} proposed using an RSA key pair with a smaller exponent and a semi-prime modulus. They based their puzzle construction on the intractability assumption in RSA, which states that computing a private exponent when the semi-prime modulus is less by multiple orders of magnitude than the public key is computationally infeasible. Given \textit{k} as a security parameter, the verification cost in the proposed scheme is reduced by a factor of $\frac{|N|}{k}$. Despite the significant cost reduction, it is still sufficiently expensive that it cannot be deployed in large-scale environments. Furthermore, their scheme does not provide fine-grained control over the difficulty level as, the gap between two subsequent difficulty levels is significantly increased compared to the previous time-lock puzzle scheme \cite{rivest1996time}. Further efficiency improvements on modular-exponentiation based puzzles are proposed in \cite{rangasamy2011efficient}. Tang and Jackmans \cite{tang2010non} introduced different verification modes for the construction phase proposed in~\cite{rivest1996time} to increase the verification efficiency and make it suitable for a server-client communication model.  

\subsubsection{Hashcash and Client puzzles}
\par Inspired by the idea of pricing functions \cite{dwork1992pricing}, Juels and Brainard \cite{juels1999client} proposed client puzzles to mitigate connection depletion attacks. The objective is to have the clients commit their resources before establishing a connection, by requiring them to solve computational puzzles for each request. Since the adversary would send a great number of connection requests, it must solve each puzzle associated with each request. Therefore, precluding attackers from overwhelming the server and allowing legitimate clients to establish a connection. Unlike previous puzzles, client puzzles do not incur high cost on the verifier, in which they are easy to construct and can be constructed in a stateless way. 
\par The server constructs the puzzle using a hash function that processes a bitstring \textit{X} of length \textit{l} as input and produces a hash image \textit{Y}. The puzzle is given as \textit{($X<k+1,l>, Y$)}, where the first \textit{k} bits of \textit{X} are hidden. Clients must perform a brute-force search for the \textit{k} missing bits that produces \textit{Y} and submit the solution with the request. The problem of solution pre-computation is addressed by embedding a time-stamp in the puzzle and requiring the client to submit the solution within a specified time interval. The difficulty is tuned by increasing/decreasing the number of bits to be searched for. To decrease the probability of guessing the solution and to have a finer difficulty adjustment, they suggest composing the puzzle of \textit{m} independent sub-puzzles, each requiring a unique \textit{k}-bit solution. This composition may increase the difficulty for an attacker in guessing the solution, but the granularity is still coarse having the difficulty level grow exponentially. Furthermore, the efficiency of the puzzle decreases as the number of sub-puzzles increases. 

\par {\bf Efficiency.} Aura et al. \cite{jakobsson1999proofs} improved the client puzzle’s efficiency by reducing its length and minimizing the number of hash calls needed for both construction and verification. They generalized the design of the puzzle to employ it in any authentication protocol. The puzzle is generated by the client, unlike the previous scheme \cite{juels1999client}, where the server generates both hash input and output. In this scheme, the client is given a nonce $N_{s}$ and is required to find a bit-string \textit{X} that if hashed together with the nonce (and other client data) produces a hash image with \textit{k} leading zero bits. For verification, the server performs a single hash call to check if the submitted solution produces \textit{k} leading zero bits. These improvements motivated further practical implementations of client puzzles. Dean and Stubblefield \cite{dean2001using} provided a compatible implementation of this scheme to protect TLS servers,  Moskowitz et al. \cite{moskowitz2008host} adapted it in the Host Identity Protocol (HIP), and Wang and Reiter \cite{wang2004mitigating} integrated it in the network layer to mitigate bandwidth-exhaustion attacks.
One additional feature that this construction offers, as proposed by Back \cite{back2002hashcash}, is that it can also be used in a non-interactive setting, where the prover chooses the challenge and the solution can be publicly verified.
\begin{figure}[h!]
\centering
\includegraphics[scale=0.25]{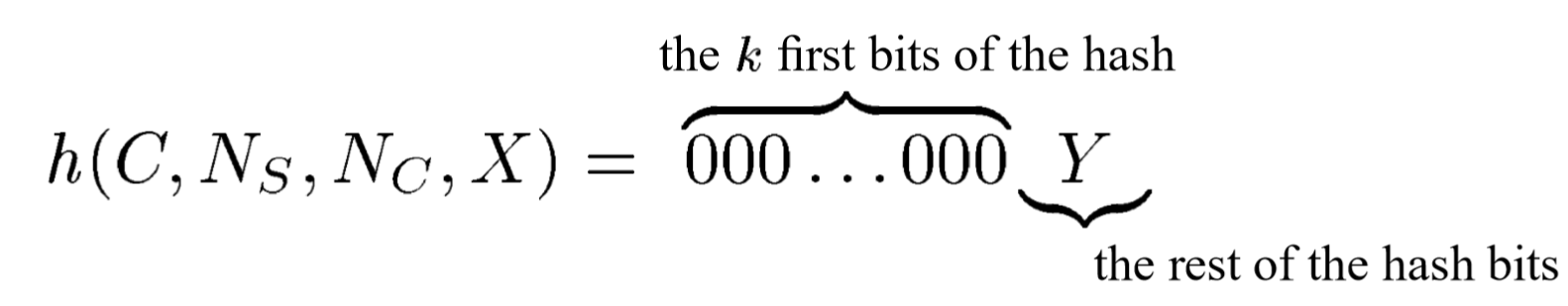}
\caption{Hashcash puzzle construction \cite{jakobsson1999proofs}}
\label{fig:hashcash}
\end{figure}

Despite the efficiency improvements, this puzzle construction, which was initially implemented by Back in the Hashcash system \cite{back2002hashcash} and currently used in Bitcoin \cite{nakamoto2008bitcoin}, provides coarser difficulty levels disabling the verifier from flexibly adjusting the difficulty according to the threat level. Furthermore, it is highly parallelizable and has a probabilistic solving cost with high variance. These drawbacks may prevent puzzle fairness and reduce the effectiveness of the puzzle. Parallelizability gives adversaries, with higher CPU-speed, a great advantage over legitimate parties, while probabilistic cost does not guarantee that the time required to solve the puzzle is similar to all clients. The search process for \textit{n}-bits might abort after the first try or after performing all of the $2^n$ tries.

\subsubsection{\revColor{Client Puzzle Variants}}\label{sub:clientvariants}
\par Further efficiency improvements were proposed and the aforementioned problems of client puzzles were addressed by several researches.  In this subsection, we categorize them based on the problems and puzzle features being discussed in each.

\par {\bf Determining the difficulty level.}
The difficulty of a puzzle determines its security and effectiveness. Setting the difficulty to a low level may reduce the workload on the adversaries which allows them to successfully launch an attack, while setting it to a high level may deter honest parties from participating in the underlying application. Given the existence of an adversary with unknown computing power, it may be difficult to properly adjust the difficulty level without increasing the cost for legitimate clients. Wang and Reiter \cite{wang2003defending} addressed this problem by using an auction mechanism. They introduced the concept of \textit{‘puzzle auctions’}, where each client is allowed to set the hardness of the puzzle it solves. The server then assigns its resources to the client that solved the hardest puzzle (the one with the highest difficulty level). They assume that a legitimate client would expend more resources than a compromised zombie client, since attackers will not increase the workload of these zombies above a certain threshold to avoid detection. They argue that this mechanism is effective, where clients can win an auction by raising their bids just above the attacker’s bid. However, it gives an unfair advantage to clients with higher computation power and, potentially, to powerful attackers. In addition, it adds more rounds to the scheme by requiring the client to submit more than one puzzle solution of increasing difficulty levels until a connection is established.

{\bf Granularity.}
Feng et al. \cite{feng2005design} addressed the coarseness problem of hash-based puzzles by providing the client with a hint along with the puzzle. The hint is a value near to the solution which reduces the brute-force search space. To solve a puzzle, the client starts at the hint and searches the range linearly for the solution. Puzzle difficulty can be adjusted finely by adjusting the accuracy of the hint. This scheme provides better control over the difficulty level, but it can only be applied interactively requiring the verifier to generate a puzzle for each client instead of just sending a challenge as in \cite{jakobsson1999proofs}. 

{\bf Reducing construction cost.}
To eliminate the computational load of generating puzzles from the verifier, Waters et al. \cite{waters2004new} proposed puzzle outsourcing mechanisms. Their goal is to protect the puzzle scheme itself from being subject to a DoS attack and reduce client delay added by the previous puzzle schemes. The puzzle, which is based on the Diffie-Hellman problem, is constructed by a third party called bastion. The construction requires modular exponentiations while the verification, which is performed by the server, requires a memory lookup and one modular exponentiation. The client is required to invert a discrete logarithm using some partial information. This information is represented as a specific range of seed values. The client performs a brute-force search for the solution seed within the specified range. Since the verifier
can control the size of the range, it can linearly adjust the difficulty of the puzzle. The client delay is reduced by allowing it to compute solutions offline and requiring it to solve a puzzle per time interval instead of per request. The per time interval requirement prevents adversaries from precomputing solutions. Although the construction is outsourced and the same bastion can generate puzzles that can be utilized by multiple servers, it is still expensive due to the modular exponentiation, making it inefficient and unscalable. In addition, the solution finding process in this scheme is similar to that in hash-based schemes, it requires performing an exhaustive search within a specific range which is a highly parallelizable task.
\par Gao \cite{gao2005efficient} suggested adding a pre-construction stage to allow the server to compute expensive operations in idle time and reuse the calculated parameters during online puzzle construction by combining them with time parameters. The author developed two trapdoor-based puzzle schemes, one is RSA-based and the other is based on the discrete logarithm problem. Both schemes require modular arithmetic calculations not only in the pre-construction stage but also in the online construction phase. As the previous scheme \cite{waters2004new}, they provide fast verification via a memory look-up and linear granularity, however, the solving process is also highly parallelizable.

\par {\bf Non-parallelizability.} Generally, non-parallelizability is important to control the timing feature of a puzzle, where the solution time should be approximately controllable by the verifier. This property is crucial to allow the verifier to appropriately set the difficulty, which is not possible due to the huge disparity in computing power between machines. Non-parallelizability is important in applications that require puzzle fairness and prevent an adversary that uses parallel computing from solving puzzles significantly faster than the expected time.
Non-parallelizable CPU-bound puzzle schemes are either based on an inherently sequential problem \cite{rivest1996time}\cite{tritilanunt2007toward} or utilize a chaining technique \cite{ma2005mitigating}\cite{groza2006chained}.
\par Time-lock puzzles \cite{rivest1996time} are based on repeated squaring which ensures non-parallelizability, however, they cannot be employed in large-scale and resource-constrained environments due to their high construction cost. Two hash-based schemes that offer non-parallelizability were proposed by Ma \cite{ma2005mitigating} and Gorza and Petrica \cite{groza2006chained}. These schemes utilize puzzle chaining technique, which requires sequential solving steps to complete the whole puzzle chain. Although they were designed to reward the client for partially solving a chain, we also discuss the possibility of using this type of construction as one puzzle that has one solution. A puzzle chain consists of a set of puzzles that can only be solved in a specific order, in which solving a puzzle requires computing the solution of the predecessor puzzle. In each puzzle chain, there is at least one puzzle that does not depend on any other puzzle. 

\par Ma \cite{ma2005mitigating} utilizes hash chain, which was initially introduced by Lamport \cite{lamport1981password}, to construct the puzzle. The goal is to provide a receiver the control over which packets to receive. This is achieved by forcing the sender to invert a hash chain, where each inverted hash value permits sending \textit{m} packets only. A \textit{‘request-to-send’} packet must be sent first by the sender before sending any further packets. The receiver then computes a hash chain ($h_{0}$, $h_{1}$,...,$h_{k}$) of length \textit{k} and responds with a puzzle given as (\textit{$h_{k}$, K}). A sender wishing to send \textit{m} packets is required to find the first predecessor hash value $h_{k-1}$ only, and is allowed to send a maximum of \textit{$m \times K$} packets by inverting the whole chain ($h_{k-1}$, $h_{k-2}$,...,$h_{0}$). To verify a single solution, the receiver is required to store the current values of $h_{k}$ and \textit{K}, then perform a single hash operation to check if the submitted hash value is equivalent to the current  $h_{k}$. The stored hash value is replaced by its predecessor after every \textit{m} packets received. Puzzle difficulty can be adjusted by varying the number of packets \textit{m} and the length of the chain \textit{K}. As a stand-alone puzzle, the construction cost is relatively high and the verification may require reconstructing the puzzle chain in some settings to avoid memory exhaustion attacks. In addition, adjusting the difficulty is not as flexible as the author claims, where inverting a chain of hash values may not always be achieved in reasonable time and setting the digest’s size to lower number of bits (\textit{16}-bits as suggested) lowers the difficulty level, which makes the process of finding a solution as easy as constructing the puzzle \cite{tritilanunt2007toward}. 

\par In Groza and Petrica \cite{groza2006chained}’s scheme, the puzzle chain is given as ($[P_{0},r_{0}],$ $[P_{1},r_{1}]...,[P_{n},r_{n}]$), where \textit{P} is the puzzle and \textit{r} is a string of random bits. The chain is constructed by first concatenating two state-dependent random values, \textit{$\rho$} and \textit{r}, then double hashing them to form the first puzzle in the chain $P_{0} = H^2(\rho_{0} || r_{0})$. The remainder of the chain is created by XORing the result of hashing the previous state-dependent values $\rho _{i-1}$ and $r_{i-1}$ with two new state-dependent values $\rho_{i}$ and $r_{i}$, then double hashing the result, $P_{i} = H^2((\rho _{i}|| r_{i}) \oplus H( \rho _{i-1}|||r_{i-1}))$. Solving the puzzle requires solving the chain in the order it was constructed, starting with $P_{0}$ then finding the $\rho_{i}$ of each $P_{i}$. Similar to the previous scheme, it is an inefficient stand-alone puzzle requiring the verifier to perform three hash operations for each puzzle in the chain for both construction and verification. Both schemes only guarantee partial non-parallelizability as, each puzzle in the chain can be solved in parallel, hence they do not actually solve the CPU-speed disparity problem.
\par To achieve non-parallelizability with cheaper construction and verification cost, Tritilanunt et al. \cite{tritilanunt2007toward} proposed a new puzzle scheme that relies on the subset sum problem, which is a variant of the knapsack problem. Given a set of objects, each has a specific weight, the prover must determine the number of objects that can be included in a fixed-size knapsack. In other words, given a set of positive integers $(a_{0},a_{1},...,a_{n}$) and a positive integer\textit{S}, find a subset of \textit{a} that sums up to \textit{S}. Solving the puzzle can be done by bruteforce which is highly parallelizable, however, the puzzle difficulty is set such that it is more efficient to apply  Lenstra’s lattice reduction algorithm \textit{LLL} \cite{lenstra1982factoring} instead. The algorithm requires recursive computations, hence it cannot be parallelized. Both construction and verification are cheap requiring one hash operation and some additions. This scheme provides non-parallelizability, deterministic computation and polynomial granularity. However, the \textit{LLL} algorithm suffers from huge memory requirements that limit its deployment in general applications. In addition, it does not solve the resource disparity problem as, according to the authors, the solving algorithm may not be executed using platforms of a lower power than PCs.


\subsubsection{Summary of CPU-bound Puzzles}
Most CPU-bound puzzle schemes are based on cryptographic hash inversions \cite{back2002hashcash, juels1999client, aura2000resistant, wang2003defending, feng2005design} or digital signature algorithms \cite{dwork1992pricing, waters2004new, gao2005efficient}. In general, hash-based puzzle schemes are more efficient where the generation and verification requires an insignificant number of cryptographic hash operations. However, they suffer from three fundamental drawbacks. The first is coarse-grained difficulty adjustment, where adjacent difficulties vary by a factor of two. This may not allow the verifier to flexibly adjust the difficulty according to the threat level, hence reducing the effectiveness of the puzzle. Although some variants, such as Hint-based \cite{feng2005design}, provide linear granularity, they can only be applied interactively. The second is parallelizability which introduces the CPU-speed disparity problem and gives powerful adversaries advantage over honest parties. Finally, the solution-finding process of these puzzles is probabilistic and its cost variance is very high that it does not guarantee fairness as some provers may get lucky and find the solution much faster than others. 
On the other hand, number-theoretic puzzles, such as time-lock puzzles \cite{rivest1996time}, provide finer granularity and low solution cost variance, but are less efficient requiring the verifier to compute large integer modular exponentiations which is unsuitable for resource-constrained environments. Considering parallelizability, both time-lock puzzles and knapsack-based puzzles ensure non-parallelizability, however as discussed previously, the former requires the verifier to perform expensive operations, while that latter suffers from huge memory requirements for solving the puzzle.

\par The lack of an efficient and non-parallelizable CPU-bound function led researchers to investigate the utilization of other resources, such as memory \cite{abadi2005moderately, dwork2003memory}, bandwidth \cite{walfish2010ddos, abliz2009guided}, and human's attention \cite{naor1996verification, von2003captcha}, which are discussed in the following sections.

\begin{center}
    \small\addtolength{\tabcolsep}{-2pt}

\begin{table*}[]
\resizebox{\textwidth}{!}{\begin{tabular}{cc|c|c|c|c|c|c|c|c|c|}
\cline{3-11}
 & \textit{\textbf{}} & \multicolumn{9}{c|}{\textit{\textbf{Properties}}} \\ \hline
\multicolumn{1}{|c|}{\textit{\textbf{Category}}} & \textit{\textbf{Puzzle Scheme}} & \textit{\textbf{\begin{tabular}[c]{@{}c@{}}Easy to\\ construct\end{tabular}}} & \textit{\textbf{\begin{tabular}[c]{@{}c@{}}Easy to\\ verify\end{tabular}}} & \textit{\textbf{Granularity}} & \textbf{\begin{tabular}[c]{@{}l@{}}Low Cost\\  Variance\end{tabular}} & \textit{\textbf{\begin{tabular}[c]{@{}c@{}}Non-\\ parallelizable\end{tabular}}} & \textit{\textbf{\begin{tabular}[c]{@{}c@{}}Non-\\ interactive\end{tabular}}} & \textit{\textbf{\begin{tabular}[c]{@{}c@{}}Publicly \\ verifiable\end{tabular}}} & \textit{\textbf{\begin{tabular}[c]{@{}c@{}}State\\ -less\end{tabular}}} & \textit{\textbf{Based on}} \\ \hline
\multicolumn{1}{|c|}{\multirow{5}{*}{\textit{\textbf{\begin{tabular}[c]{@{}c@{}}Non-Hash \\ Based\end{tabular}}}}} & Merkle \cite{merkle1978secure} & \xmark & \cmark & Exponential & \xmark & \xmark & \xmark & \xmark & \xmark & \begin{tabular}[c]{@{}c@{}}Symmetric \\ Crypto.\end{tabular} \\ \cline{2-11} 

\multicolumn{1}{|c|}{} & Pricing functions \cite{dwork1992pricing} & \xmark & \cmark & Exponential & \xmark & \xmark & \cmark & \xmark & \xmark & \begin{tabular}[c]{@{}c@{}}Digital\\  Signatures\end{tabular} \\ \cline{2-11} 

\multicolumn{1}{|c|}{} & Time-lock \cite{rivest1996time} & \xmark & \cmark & Linear & \cmark & \cmark & \xmark & \cmark & \xmark & \begin{tabular}[c]{@{}c@{}}Repeated \\ Squaring\end{tabular} \\ \cline{2-11} 

\multicolumn{1}{|c|}{} & Subset-Sum \cite{tritilanunt2007toward} & \cmark & \cmark & Polynomial & \cmark & \xmark & \xmark & \cmark & \cmark & \begin{tabular}[c]{@{}c@{}}Subset-Sum \\ problem\end{tabular} \\ \cline{2-11} 

\hline
\multicolumn{1}{|c|}{\multirow{4}{*}{\textit{\textbf{Hash-Based}}}} & Client Puzzles \cite{juels1999client} & \cmark & \cmark & Exponential & \xmark & \xmark & \xmark & \xmark & \cmark & \begin{tabular}[c]{@{}c@{}}Multiple \\ Hash Inversion\end{tabular} \\ \cline{2-11} 

\multicolumn{1}{|c|}{} & Hashcash \cite{back2002hashcash} & \cmark & \cmark & Exponential & \xmark & \xmark & \cmark & \cmark & \cmark & \begin{tabular}[c]{@{}c@{}}Single \\ Hash Inversion\end{tabular} \\ \cline{2-11} 

\multicolumn{1}{|c|}{} & Hint-based \cite{feng2005design} & \cmark & \cmark & Linear & \xmark & \xmark & \xmark & \xmark & \xmark & \begin{tabular}[c]{@{}c@{}}Single \\ Hash Inversion\end{tabular} \\ \cline{2-11} 

\multicolumn{1}{|c|}{} & Hash-chain \cite{ma2005mitigating}\cite{groza2006chained} & \xmark & \cmark & Linear & \xmark & partial & \xmark & \xmark & \xmark & \begin{tabular}[c]{@{}c@{}}Hash chain\\  reversal\end{tabular} \\ \hline
\\
\end{tabular}}

\centering
\caption{Summary of CPU-bound schemes.}
\label{table:cpu}
\vspace{-0.8cm}
\end{table*}
\end{center}

\subsection{\textbf{Memory-Bound and Memory-Hard Schemes}} \label{memory-sec}
\par Using memory-intensive computations in a puzzle scheme has been proposed by several works \cite{abadi2005moderately, percival2009stronger, dziembowski2015proofs, ateniese2014proofs}, with different notions, to provide both equitable computation and resistance against specialized hardware, such as GPUs and ASICs. The solution cost of the puzzle is measured either by the number of accesses to the main memory or the amount of memory space rather than the number of computation operations. 

\par \textbf{\textit{Memory-bound}} puzzle schemes require a significant number of memory accesses to be solved. The solution time is bound by the memory latency, in which the complexity is measured by the number of cache misses and not the actual amount of memory being employed. The adversary’s goal in such a scheme is to decrease the number of memory accesses by either benefiting from cache or performing intensive computations instead. On the other hand, \textbf{\textit{memory-hard}} puzzle schemes require a significant amount of memory space to solve. The complexity is measured by the number of memory locations being used for a given number of operations. The adversary’s goal in such a scheme is to use less memory space by trading it for time or extra computations. 

\par A memory-bound function may be considered as memory-hard in the sense that the amount of memory required is greater than the cache's size. While a memory-hard function is memory-bound only if the locality is good in its memory access pattern such that it results in a certain number of cache misses \cite{ren2017bandwidth}. 


\subsubsection{Easy-to-compute functions}
A well-known problem with CPU-bound puzzle schemes is the significant variations in the amount of computing power available to provers. Abadi et al. \cite{abadi2005moderately} addressed this problem and were the first to introduce the notion of memory-bound puzzles. The goal is to construct a puzzle that latest systems can compute at a similar speed by relying on memory latency instead of CPU-speed. They proposed a memory-bound puzzle that consists of an easy-to-compute pseudo-random function \textit{F()}, which its inverse $F^{-1}()$ requires more time-consuming computations than accessing the memory (i.e. inverting it can be done more efficiently via the space-time tradeoff). The verifier selects an integer $x_{0}$ from the domain $[0…(2^{n}-1)]$ and computes $x_{i+1} = F(x_{i})\oplus i$, where $0 < i < k$. The puzzle is given as $x_{k}$ and a checksum of the sequence $(x_{0}, x_{1},...,x_{k})$. Solving the puzzle requires constructing a table for $F^{-1}()$ and working backwards from  $x_{k}$ to find a pre-image $x_{0}'$, such that the checksum of the path from $x_{0}'$ to $x_{k}$ matches that of the challenge. Since there exist several pre-images that lead to multiple paths, the solver is forced to explore a tree of pre-images that has a depth \textit{k}, root $x_{k}$ and a total size of $O(k^{2})$. The parameters \textit{n} and \textit{k}, are chosen carefully to ensure that the table cannot be stored in cache, thus in the best case scenario, the number of cache misses required is also $O(k^{2})$. Verification requires \textit{k} forward computations of \textit{F()}, which increases exponentially as the CPU-cost of \textit{F()} decreases (i.e. the processing speed of the current machines increases according to Moore’s Law). 
\par They also discuss a variant of this scheme that can be applied in a non-interactive setting. The challenge is not presented by the verifier, rather produced by a pseudo-random generator using some application related data, such as a message in combating spam, as the seed.
\par The main drawback of the scheme proposed in \cite{abadi2005moderately} is the existence of a time-space tradeoff for inverting the function \textit{F()}, which may allow adversaries to circumvent the scheme using higher computation power and only rarely accessing the memory. The puzzle complexity is highly affected by current processing speeds, in which it requires more memory and higher verification cost to keep pace with Moore’s law. Consequently, its efficiency decreases which hinders its large-scale deployment and practical implementation. In addition, the work ratio between the two parties is quadratic and cannot be increased since deeper trees allow the solver to benefit from cache and invert several values in the cost of just one memory access. 

\subsubsection{MBound}
\par Dwork et al.\cite{dwork2003memory} argued that easy-to-compute functions \cite{abadi2005moderately} may be solved with very few memory accesses which diminishes the memory latency effect and hence do not solve the problem. They further explored memory-bound functions and defined a class of functions based on \textit{“pointer-chasing”} in a large random table \textit{T}. The table \textit{T} is of a fixed size that is approximately double the size of the largest current existing cache and is shared between the two parties. The solver is forced to perform random walks through \textit{T} to find a path with specific characteristics. A walk is a path that requires making a series of sequential random accesses to \textit{T}, where contents of the current accessed location determine the subsequent location to be accessed. Solving the puzzle is done by executing an algorithm (called \textit{MBound}) that terminates successfully if after \textit{l} walking steps the last \textit{n}-bits of the hash output are zeros. The algorithm is executed $k$ times until a successful path is found. The prover then submits the solution given as the hash-output along with the trial number $k$, which identifies the correct path. The verifier then explores the identified path \textit{k} by performing \textit{l} memory accesses to \textit{T} and checks that the submitted hash value is correct and ends with $n$ zeros. 
\par Using the random oracle model the authors prove a lower bound of $\Omega(2^{n}\cdot l)$ on the number of amortized memory accesses that an adversary must expend per puzzle. The number of walks to be performed by the solver is $2^{n}$, thus the total expected cost for computing the puzzle is $2^{n}\cdot l$ cache misses, while the verification cost is \textit{l} cache misses.  

\par On a follow-up work, Dwork et al.\cite{dwork2005pebbling} addressed the high communication complexity required for distributing and updating the incompressible large table by allowing the table to be constructed using graph pebbling. They observed that the table construction must also be memory-bound in order to prevent adversaries from exploiting the compact description of \textit{T} and produce elements in cache whenever they are needed, only rarely accessing the memory. 

{\bf Graph Pebbling.} Pebbling is described as a game played on a directed acyclic graph (DAG). Finishing the game requires pebbling all output vertices (nodes with no children) of the graph. A non-input vertex (a node with parents) can only be pebbled when all of its predecessor vertices (parents) are pebbled. A pebble can be removed from the graph at any time. The player has a number of pebbles and its goal is to place the pebbles on the output vertices efficiently, by using few moves and having few pebbles on the graph at any time. Graph pebbling is used by several papers to model memory-bounded computations, where the time-space tradeoff is obtained by showing that there is no optimal pebbling strategy that uses a few simultaneous pebbles. Generally, each pebble represents the output of a computation and performing a specific computation using previously computed output is represented as the placing of a pebble. A more detailed description of pebble games is presented in \cite{liu2017red}.
\par In this scheme, a pebble corresponds to a  label of \textit{n}-bits and placing a pebble corresponds to labeling a vertex by calling a hash function and storing the newly computed value in cache. Constructing the table requires pebbling the DAG \textit{D}, where the labels of the output vertices are the elements of table \textit{T}. The DAG has \textit{N} input vertices (\textit{numbered 1,2,...N}), \textit{N} output vertices (\textit{numbered N+1, N+2,..,2N}) and a constant indegree. The size of the graph is $O(n|T|log|T|)$ and consists of a stack of \textit{N}-superconcentrators, which provide sharp tradeoffs having the time required to pebble the output vertices with less than $N$ pebbles at least exponential in the depth of the graph. Therefore, the time required to pebble the outputs is superpolynomial in $|T|$ (the number of elements in \textit{T}). 
Table construction is done by both parties only once, the table is then stored in main memory to execute the \textit{MBound} algorithm \cite{dwork2003memory}. This reduces communication cost, however, the requirement of constructing a table incurs a relatively high cost on the verifier and eliminates the easy-to-construct property.

There are two main drawbacks of the aforementioned memory-bound puzzles that hinder their deployment. First, both verification and solving costs increase to accommodate to the current cache size, which reduces the puzzle’s efficiency and disables legitimate parties with low memory resources from participating in the scheme. Second, the difficulty level cannot be adjusted flexibly, in which it is constrained by the requirement of fine-tuning several parameters depending on the system’s cache and memory configurations. 

\subsubsection{Pattern Database}
\par Doshi et al. \cite{doshi2006efficient} noticed that the memory accesses described in prior memory-bound puzzle schemes are not associated with a known problem, hence lack the algorithmic foundation that allows flexible tuning of difficulty parameters. They were the first to propose deriving memory-bound puzzles from heuristic search methods.

\par The proposed construction scheme is based on a heuristic search technique using pattern databases for the Sliding Tile problem \cite{loyd1959mathematical}. The Sliding Tile problem requires finding a path (using only the following moves: \textit{left, right, up, down}) to slide the tiles on a grid from the initial state to the target state.

Solving the problem can be done efficiently using existing computational algorithms, such as the \textit{A*} algorithm together with the Manhattan Distance heuristic. However, using a memory-based heuristic is more efficient, where the precise distance from the initial state to the abstract target state is computed and stored in a look-up table called \textit{pattern database}. 
\par Given a pool of target states ($G_{0}...G_{n}$), the prover precomputes the pattern database that corresponds to these states offline. The verifier constructs the puzzle by randomly choosing \textit{f} target states from the pool and performing \textit{d} moves at random to each target $G_{i}$ in order to create \textit{f} initial states. The verifier then computes a message authentication code (MAC) over the performed moves and stores it in memory. The difficulty of the puzzle can be tuned by increasing or decreasing the number of target states \textit{f} and the number of moves \textit{d}. The puzzle is given as the \textit{f} target states, the corresponding initial states, and checksums $C_{j}$, $1\leq j \leq d$, where each checksum is computed over the \textit{j}th move of each state. 
\par The prover then uses the computed pattern database to complete a guided search for every target state. All initial states are solved simultaneously as, the prover must first deduce the right set of moves for a given difficulty level \textit{i}, $1\leq i \leq d$, such that the checksum of the corresponding moves matches $C_{j}$. Since different states are stored in different parts of the memory, forcing the prover to search for multiple states simultaneously results in cache misses hence ensures main-memory access.  
Once all the target states are reached, the prover submits the \textit{d} moves performed on the \textit{f} initial states. The verifier then checks that the moves are correct by computing the MAC of these moves and comparing it to the one stored. 
\par Although the scheme provides better flexibility in tuning the puzzle difficulty, it is inefficient in terms of both construction and communication costs. For each puzzle, the verifier is required to perform $d*f$ moves and compute $d$ checksums, while communication involves transmitting two sets of states (both initial and target) along with $d$ checksums and a MAC.


\subsubsection{Scrypt}
In 2009, Percival \cite{percival2009stronger} proposed Scrypt and introduced the concept of memory-hard functions that entail significant amount of memory to evaluate and require a large number of computations if less memory is utilized. Scrypt is a memory-hard key derivation function used for password hashing to increase the cost of brute-force dictionary attacks, whereby the attacker iterates through a number of likely passwords and apply the function to each password guess. The designing goal of the scheme is to reduce the advantage gained by adversaries who use custom-designed parallel circuits, while maintaining low per-evaluation cost of the honest user. This is achieved by requiring an amount of memory that is approximately proportional to the number of operational steps performed to evaluate the function. 

\par The scheme consists of several memory-hard functions, we briefly describe the construction of ROMix which constitutes the core of Scrypt. The basic idea of the algorithm, as described by Percival, is to sequentially compute a large number of random values and then access each value randomly to ensure that they are all stored in RAM. Given a $k$-bit input value $B$, a chain of \textit{N} input values ($X_{0},...,X_{N-1}$) is computed as follows: \begin{itemize}
    \item $X_{0} = B$ and $X_{i} = H(X_{i-1})$ for $i = 1,..,N-1$, where \textit{H} is a hash function.
\end{itemize}
A chain of \textit{N} output values ($V_{0},...,V_{n}$) is then computed as follows: 
\begin{itemize}
    \item $V_{0} = H(X_{N-1}) $, $V_{i} = H(V_{i-1}\oplus X_{V_{i-1}\ mod\ N})$ for $i = 1,..,N$, where $V_{N}$ is the final output of the function.
\end{itemize}

The default strategy of computing Scrypt is to sequentially compute each $X_{i}$ in the input chain and store it in the memory, and then compute each $V_{i}$ of the output chain and fetch each $X_{i}$ from memory as needed in order to produce the final output $V_{N}$. Two recent works show that Scrypt provides almost optimal resistance against ASICs from both the area \cite{alwen2017scrypt} and energy \cite{ren2017bandwidth} aspects. However, this is only achieved if the memory requirement of the scheme is large enough, which consequently incurs high construction and verification costs. 

\par Scrypt is used as a pricing puzzle by several cryptocurrencies including Litecoin and Dogecoin, where the amount of memory is reduced in order to provide faster verification. To append a block to the chain, the miner is required to find a nonce \textit{n} that if hashed along with the block's header \textit{B} using Scrypt, produces a final output $V_{N}$ that is less than a specific target value \textit{T}. The solution is given as (\textit{B}, \textit{n}, $V_{N}$). Verification requires a single call to Scrypt and a single comparison operation to check the output value ($V_{N}$) against the target value \textit{T}. The memory hardness of Scrypt can be set by increasing or decreasing the CPU/memory cost parameter \textit{N}, while puzzle difficulty is adjusted by tuning the target value \textit{T}. 

\par Although a work gap between the two parties is maintained by having the prover perform multiple Scrypt calls to find the solution while only a single call is required for verification, the scheme is symmetric in terms of memory requirements. Furthermore, its adaptation in Litecoin cryptocurrency is not ASIC-resistant having ASIC mining rigs hundreds of times more efficient. 

\begin{center}
    \small\addtolength{\tabcolsep}{-3pt}
\begin{table*}[]
\resizebox{\textwidth}{!}{\begin{tabular}{cc|c|c|c|c|c|c|c|c|c|}
\cline{3-11}
 & \textit{\textbf{}} & \multicolumn{9}{c|}{\textit{\textbf{Properties}}} \\ \hline
\multicolumn{1}{|c|}{\textit{\textbf{Category}}} & \textit{\textbf{Puzzle Scheme}} & \textit{\textbf{\begin{tabular}[c]{@{}c@{}}Easy to\\ construct\end{tabular}}} & \textit{\textbf{\begin{tabular}[c]{@{}c@{}}Memory-less\\ verification\end{tabular}}} & { \textit{\textbf{Granularity}}} & \multicolumn{1}{l|}{\textbf{\begin{tabular}[c]{@{}l@{}}Low Cost\\  Variance\end{tabular}}} & \textit{\textbf{\begin{tabular}[c]{@{}c@{}}Non-\\ parallelizable\end{tabular}}} & \textit{\textbf{\begin{tabular}[c]{@{}c@{}}Non-\\ interactive\end{tabular}}} & \textit{\textbf{\begin{tabular}[c]{@{}c@{}}Publicly \\ verifiable\end{tabular}}} & \textit{\textbf{\begin{tabular}[c]{@{}c@{}}State\\ -less\end{tabular}}} & \textit{\textbf{Based on}} \\ \hline
\multicolumn{1}{|c|}{\multirow{5}{*}{\textit{\textbf{\begin{tabular}[c]{@{}c@{}}Heuristic \\ Search\end{tabular}}}}} & 
\begin{tabular}[c]{@{}c@{}}Easy-to-compute \\ functions \cite{abadi2005moderately}\end{tabular} & \cmark & \cmark & polynomial & \xmark & partial & \xmark & \xmark & \xmark & \begin{tabular}[c]{@{}c@{}}Depth-first \\ search\end{tabular} \\ \cline{2-11}
\multicolumn{1}{|c|}{} & \begin{tabular}[c]{@{}c@{}}Pattern \\ Database \cite{doshi2006efficient}\end{tabular} & \xmark & \cmark & polynomial & \xmark & partial & \xmark & \xmark & \cmark & \begin{tabular}[c]{@{}c@{}}Sliding Tile \\ Problem\end{tabular} \\  \cline{2-11}
\multicolumn{1}{|c|}{} & Cuckoo Cycle \cite{tromp2015cuckoo} & \cmark & \cmark & exponential &  \xmark & \xmark & \cmark & \cmark & \cmark & \begin{tabular}[c]{@{}c@{}}Graph\\ Cycle Finding\end{tabular} \\ \cline{2-11} 
\multicolumn{1}{|c|}{} & Equihash \cite{biryukov2017equihash} & \cmark &  \cmark & exponential & \xmark & partial & \cmark & \cmark & \cmark & \begin{tabular}[c]{@{}c@{}}k-XOR\\ Birthday Problem\end{tabular} \\ \cline{2-11} \hline
\multicolumn{1}{|c|}{\multirow{6}{*}{\textit{\textbf{\begin{tabular}[c]{@{}c@{}}Graph\\ Pebbling\end{tabular}}}}} & \begin{tabular}[c]{@{}c@{}}Compact-\\ MBound \cite{dwork2005pebbling}\end{tabular} &  \xmark  & \xmark & exponential & \xmark & partial & \cmark & \xmark & \xmark & \begin{tabular}[c]{@{}c@{}}Graph\\ Pebbling\end{tabular} \\ \cline{2-11} 
\multicolumn{1}{|c|}{} & \begin{tabular}[c]{@{}c@{}}Litecoin-\\ (Scrypt)\cite{lee2011litecoin} \end{tabular}  & \cmark & \cmark & exponential & \xmark & \cmark & \cmark & \cmark & \cmark & \begin{tabular}[c]{@{}c@{}}Graph\\ Pebbling Reduction\end{tabular} \\ \cline{2-11} 
\multicolumn{1}{|c|}{} & \begin{tabular}[c]{@{}c@{}}MTP-\\ Argon2d \cite{biryukov2016egalitarian}\end{tabular} & \cmark & \cmark & exponential & \xmark & partial & \cmark & \cmark & \cmark & \begin{tabular}[c]{@{}c@{}}Graph\\ Pebbling Reduction\end{tabular} \\ \cline{2-11} 
\multicolumn{1}{|c|}{} & Proof of space \cite{dziembowski2015proofs} & \xmark  & \cmark & linear  & \xmark & partial  & \xmark  & \xmark  & \xmark  &  \begin{tabular}[c]{@{}c@{}}Graph\\ Pebbling\end{tabular} \\ \cline{2-11} \hline
\\
\end{tabular}}
\centering
\caption{Summary of Memory-bound schemes.}
\label{table:memory}
\end{table*}
\end{center}

\subsubsection{Cuckoo Cycle}
Tromp \cite{tromp2015cuckoo} proposed a memory-bound puzzle scheme based on finding constant sized cyclic subgraphs in a pseudo-random graph. The designing goal is to make mining cryptocurrencies on commodity hardware cost-effective by relying on memory-latency instead of computation speed in the evaluation of the puzzle. 
The puzzle is constructed by generating a directed bipartite graph with \textit{N} vertices and \textit{M} edges from a given set of input nonces using Cuckoo hashing \cite{pagh2004cuckoo}. The prover is required to find a set of \textit{L} nonces whose corresponding edges form an \textit{L}-cycle in the graph and whose hash digest is less than a given target hash value \textit{T}, where $0 \leq T \leq 2^{256}$. The memory-hardness of the scheme is determined by the ratio $M/N$ and the cycle length \textit{L}, while the difficulty of the puzzle is adjusted by tuning the target hash value \textit{T}. The verification cost is linear in \textit{L} and independent of both \textit{M} and \textit{N}.

\subsubsection{Proof of Space}
Dziembowski et al. \cite{dziembowski2015proofs} and Ateniese et al. \cite{ateniese2014proofs} introduced the notion of \textit{proof of space} (PoS) independently, each with a different definition and security guarantees. Generally, PoS is an interactive memory-hard puzzle scheme where the solution serves as a compact proof that the prover did possess and dedicate a significant amount of space. Unlike memory-bound schemes \cite{abadi2005moderately, dwork2003memory, tromp2015cuckoo}, PoS schemes rely on memory space instead of memory latency in the evaluation of the puzzle. Both PoS construction schemes are based on directed acyclic graphs (DAGs) that have high pebbling complexity and use the Merkle hash tree to enable efficient verification. 

\par Given a challenge nonce \textit{id}, the prover is required to build a graph G = (V, E) of \textit{N} vertices and store the label of each vertex $v$ computed as $l_v = H_{id}(v,\ H_{id}(l_{p1},..,l_{pd}))$, where $H_{id}$ is a hash function that depends on the identifier \textit{id} and $l_{p1},...,l_{pd}$ are the labels of $v$'s predecessors (parents). The identifier is used to ensure that the same dedicated space cannot be leveraged for more than one proof. Once the labels are computed, the prover commits to these labels by constructing a Merkle hash tree $\tau = \tau^{H}(l_1,...l_N)$ and submitting the computed root of the tree, $\Phi$, to the verifier. At this point, the verifier checks the consistency (correctness) of the root $\Phi$ by asking the prover to open the labels of \textit{\textbf{c}} vertices and their predecessors. The opening of a label $l_{v_{c_{i}}}$ is the path from the root $\Phi$ to the leaf associated to vertex \textit{\textbf{v}} in the Merkle tree $\tau$. Given the labels of all the predecessors of vertex \textit{\textbf{v}}, the verifier can check if the label $l_{v_{c_{i}}}$ is correctly computed as described above. The solution is accepted only if all \textit{\textbf{c}} openings and labels are computed correctly.  To label a vertex \textit{v} in graph G, the prover must compute and save on his memory the label values of all the predecessors $l_{p1},..,l_{pd}$ of the vertex. The labeling of the graph presents a proof that the prover has handled at least \textit{N} space. 
\par The definition of PoS introduced in \cite{dziembowski2015proofs} extends that of Ateniese et al. \cite{ateniese2014proofs} by including an additional phase, called \textit{execution phase}, that allows the verifier to repeatedly challenge the prover and check if it is still dedicating the specified amount of space. The purpose of the additional phase is to allow honest provers to respond to the repeated challenges by accessing the \textit{N} space, produced in the initialization phase, while using little computation hence addressing the high energy consumption drawback of CPU-bound puzzles. This formation of PoS is also referred to as proof of persistent space by Ren and Devadas \cite{ren2016proof}.   

\par Although these construction schemes provide strong security guarantees by forcing a cheating prover to either expend $O(N)$ space or time in order to produce an acceptable proof, they are both based on superconcentrators which are relatively slow \cite{biryukov2016egalitarian}. 

\subsubsection{Equihash}
Biryukov and Khovratovich \cite{biryukov2017equihash} observed that any determined NP-complete problem can be used to design a memory-bound puzzle with tunable parameters, where its memory hardness is determined by the time-memory tradeoffs of the best known algorithms. They proposed Equihash, a proof-of-work scheme based on the k-dimensional generalized birthday problem \cite{wagner2002generalized}. In the generalized birthday problem, there are \textit{k} sets of \textit{n}-bit strings and the goal is to find \textit{k} strings that XOR to zero. In Equihash, the \textit{k} strings are generated randomly using the hash function \textit{H}. The optimal solution algorithm to this problem has a time and space complexity of $O(2^{\frac {n}{k+1}})$ \cite{wagner2002generalized}), hence it is a memory-intensive algorithm. 
In addition, using $1/q$ less memory results in $O(q^{\frac {k}{2}})$ times more calls to the hash function, which limits the computation advantage of parallelization to the amount of memory bandwidth available. The verification requires performing $2^k$ hashs and XORs.
%

\subsubsection{MTP-Argon2d}
Biryukov et al. \cite{biryukov2016egalitarian} proposed a \textit{non-interactive} memory-hard puzzle scheme, called Merkle Tree Proof (MTP), based on the memory-hard function Argon2d \cite{biryukov2016argon2}. They describe different instantiation settings of the scheme to be applied in crypto-currency, time-release cryptography and disk encryption. Similar to PoS schemes \cite{dziembowski2015proofs, ateniese2014proofs}, this scheme leverages a Merkle hash tree construction over an array of memory segments to allow fast and memory-less verification. 
\par Given the challenge \textit{I}, the prover constructs the puzzle by generating \textit{M} segments of memory $B[1], B[2],...,B[M]$ using Argon2d from the given challenge. It then constructs a Merkle hash tree over the \textit{M} segments committing to the segments' values. The root of the Merkle tree is declared as $\Phi$. Each memory segment $B[j]$ is an output of of Argon2d's compression function \textit{F}, which processes the preceding segment $B[j-1]$ and $B[\phi(j)]$, where $\phi(j)$ is a data-dependent indexing function that produces a memory segment index ranging in $[1...j)$. 

The prover solves the puzzle by accessing \textit{L} memory segments of $B[1], B[2],...,B[M]$ pseudorandomly in the sequence $B[j_1], B[j_2],.., B[j_{L-1}], B[j_{L}]$ to find a nonce \textit{N} that if hashed with the root $\Phi$ and the segments $B[j_i]$ produces a hash output $Y_L$ with \textit{d} trailing zeros (where $j_i$ is determined from the nonce \textit{N}, the root $\Phi$, and the values of the preceding segments in the sequence). The solution is given as ($\Phi, N, \mathscr{L}$), where $\mathscr{L}$ is the opening of $2*L$ memory segments {$B[j_i-1], B[\phi(j_i)]$}. The verifier then validates the openings $\mathscr{L}$ and regenerates all {$B[j_i] = F(B[j_i-1], B[\phi(j_i)])$} to verify that $Y_L$ has \textit{d} trailing zeros. The memory hardness of the scheme is highly dependent on the hardness of Argon2d and can be adjusted by tuning the parameters. The difficulty of the puzzle is tuned by increasing or decreasing the number of trailing zeros \textit{d}.

\subsection{\textbf{Bandwidth and Network Bound Schemes}}\label{sub:Band_netw_bound_schemes}
\par Walfish et al. \cite{walfish2010ddos} were the first to suggest using bandwidth as a currency to pay for a service in order to mitigate DoS attacks. They did not propose a puzzle scheme, but introduced the idea of using bandwidth resources to weaken powerful attackers with higher CPU and memory resources than legitimate clients. The proposed system, called \textit{Speak-up}, crowds out attackers by encouraging clients to send higher volumes of traffic for their legitimate requests to be served first. The authors assume that attackers are already dedicating the highest amount of upload bandwidth resources to perform the DoS attack which prevents them from reacting to the encouragement.
\par In what follows, we describe the different puzzle construction schemes found in the literature that rely either on bandwidth resources or network latency in the evaluation of the puzzle.

\subsubsection{Guided Tour Puzzles}
\par Abliz and Tznati \cite{abliz2009guided} introduced the idea of network-bound puzzles to overcome the shortcomings of previous client puzzle schemes, which include parallelizability and computational disparities among clients. Their scheme requires the client to collect tokens from \textit{tour guides }(a pre-specified set of nodes) in order to be able to solve the puzzle and have their query processed by the service provider. The tour guides are used to introduce a delay between client requests. The authors suggest network latency as a solution to eliminate the disparity in the amount of resources between a powerful adversary and a legitimate client. Non-parallelizability is achieved by the random selection of the next tour guide to be visited at each tour guide stop (hence the name guided tour) and by requiring the client to submit the hash values from the previous tour guides at each stop. The difficulty is tuned by increasing or decreasing the tour length (number of tour guides) by one which provides linear puzzle granularity. 

\subsubsection{Bandwidth Puzzles}
Reiter et al. \cite{reiter2009making} proposed a bandwidth-bound puzzle scheme to validate that a peer did actually expend a certain amount of communication resources and relayed data to peers in a P2P content-distribution system. The main goal is to prevent colluding attackers from earning rewards by claiming that they have exchanged data to each other without actually transferring any data. \revColor{If a file sharing P2P network does not implement a robust incentive mechanism, some peers could trick the system by claiming to have transferred a certain amount of files to other peers, without having done so, causing the entire system to fail~\cite{zhang2012new}. 
For example, if Alice and Bob are friends, Alice may claim to have transferred a certain amount of data to Bob. When the P2P system asks Bob about this transaction, Bob will confirm that he has received the data because he is Alice's friend. In this way, Alice will get improper credits. 
In the proposed scheme, the P2P network includes a central credit management entity, the verifier. When the verifier suspects collusion between two or more clients, he will simultaneously send a puzzle to the suspected peers, who then become the provers of the bandwidth puzzle scheme. The puzzle must have at least two main properties: (i) the solution stage must take time; and, (ii) the puzzle can only be solved if the prover owns the file in question. When the verifier suspects collusion between Alice and Bob, he will simultaneously send two different puzzles, one to Alice and one to Bob, asking the solution within a certain time threshold. If Alice has not shared the file with Bob, Alice can solve her puzzle and send the solution to the verifier within the threshold, while Bob cannot do the same, since he does not have the file. 
At this point, Bob could send his puzzle to Alice, asking her to solve it. Although Alice is able to solve Bob's puzzle, she is unable to solve two puzzles within the threshold. For this reason, At least one between Alice or Bob will fail in replying to the verifier with the correct solution, revealing the collusion.} Their puzzle scheme can be considered relevant to proofs of data possession mechanisms as the solution depends on the possession of the content. Finding the solution is relatively easy for provers who possesses the content, while more difficult for those who do not. Each puzzle is composed of a hash function and a collection of index-sets, where each set contains k random content indices. The verifier constructs a puzzle by hashing the content bits which are indexed by a pseudorandomly selected index-set and attached together in a certain order. Finding the solution of the puzzle requires determining which of the index-sets is the pre-image of the hash. The puzzles are issued simultaneously to a group of provers and must be solved within a specific time. This prevents a content-holder from colluding with others by solving their puzzles, since he can only solve one puzzle at a time. On a follow up work, Zhang \cite{zhang2012new} studied the effectiveness of the proposed scheme \cite{reiter2009making} and provided a lower bound on the number of content bits attackers should possess to be able to defeat the scheme with a definite probability. \revColor{This lower bound could be used to properly set the puzzle's parameters in real-world systems to improve the overall security.}

\section{Further Developments}\label{subsec:further_dev}
After discussing the main approaches used in designing the different types of puzzles, it is possible to draw a high-level view of the state-of-the-art in this field. The limitations highlighted in Section~\ref{schemes} could be used to identify future developments, driven by the need to improve the performance/effectiveness of the puzzles. Moreover, brand-new properties required by the adoption of new technologies could also lead to further advancement in puzzle construction schemes. 

\subsection{Edge/Fog Architecture} The microservice architecture, enabled by several virtualization techniques such as containers and unikernels, is increasingly used in the cloud environment, as well as in the edge/fog architecture~\cite{8812214}. In this context, bandwidth/network bound puzzle schemes could be very important to mitigate one of the major security concerns, the DDoS attack problem, which aims to destroy the availability of services. Unfortunately, existing schemes described in Section~\ref{sub:Band_netw_bound_schemes} are not feasible for this specific scenario, due to the high workload required at the server-side to generate the puzzle. For this reason, further research efforts are required to design more efficient schemes with low construction and verification costs, 
 suitable for this kind of scenario. Moreover, bandwidth/network bound puzzle schemes are normally used too high in the TCP/IP protocol stack (i.e. application layer), leaving the underlying layers exposed to DDoS attacks. By moving the application of the puzzles to the lower layer, it may be possible to block DDoS attacks in advance, also protecting the upper layers. 
 
\subsection{Resource Constraint Devices} The massive adoption of IoT networks has significantly changed the architecture of end-user applications, designed to minimize latency in client-server communications and works on devices with limited resources. Many research efforts have been made to ensure the portability of existing puzzle technologies on this new architecture, but some use-cases, such as blockchain-based applications, still require several advancements. The high amount of electricity required to solve some CPU-bound puzzles used in different cryptocurrency schemes has triggered a heavy debate~\cite{DEVRIES2018801}, that highlighted the weaknesses of this type of puzzles. Consequently, several new memory-bound schemes have been proposed and used as proof-of-work in some cryptocurrencies~\cite{8695663}, to overcome these issues. One of the most successful proposals in this context is CryptoNote~\cite{CryptoNote}, a memory-bound scheme designed to ensure the fairness property of the puzzle. This scheme has been used as a proof of work in several cryptocurrencies, in an attempt to make CPU-based mining equally efficient compared to the GPU-based one, limiting the advantages of using ASICs\footnote{Application Specific Integrated Circuits: hardware systems specifically designed for crypto-mining activities.} hardware for crypto-mining activities. However, the CryptoNote scheme has not been sufficiently discussed in the literature. For this reason, given the increasing attention it is receiving, important research efforts are needed to verify the soundness of the mathematical properties underlying this scheme, as well as the resistance against the circumvent memory attack. Moreover, some of its properties make the cryptocurrencies based on this scheme appealing to malicious actors, because also resource constraint devices can be used for mining activities~\cite{8514837}. Consequently, the need to modify the characteristics of the puzzles in order to prevent illegal behavior, as well as to develop valid countermeasures arises, opening new interesting research perspectives.

\subsection{Anonymous Networks} Anonymous communications have been introduced to increase the users' privacy within the shared public network environment. Their aim is to provide anonymity between users, apart from content privacy and integrity, ensured by other technologies~\cite{REN2010420}. Examples of anonymous networks are Virtual Private Networks (VPNs), Onion routing, web MIXes, peer-to-peer anonymous communications systems, and possibly others. These systems, apart from other attacks that compromise the anonymity, are vulnerable to DDoS attacks that aim to disrupt the service.\\
Acronym of The Onion Router, the Tor network~\cite{dingledine2004tor} not only allows users to be anonymous to the website, but also the website to be anonymous to the users. This requires the establishment of a circuit that makes use of different levels of encryption. Beyond the privacy problems that users can still be affected to~\cite{la2018time, sun2015raptor}, the Tor network can be subjected to different denial of service attacks. An example of this type of attack was discussed in~\cite{ barbera2013cellflood}, that proposes CellFlood, a brand-new denial of service attack targeting Tor routers that impacts the router ability to create a new circuit. The same authors, in the same paper, have shown how a puzzle can be used even in this context, since they allow Tor routers under attack to slow down the attacking hosts, which maintains their ability to manage legitimate client requests. However, the authors used a CPU-bound puzzle, that does not meet the key requirements of the peer-to-peer network puzzles, discussed in Section~\ref{subsub-p2p}, most of which are also valid for the Tor Network. In this context, further research efforts are needed to design new puzzle schemes feasible for the anonymous network environment. Bandwidth/network bound puzzle schemes could be considered in this scenario, with the aim of improving the efficiency and fairness of the overall architecture.

\revColor{\subsection{New Generation of Mobile Networks}\label{sub:5G}
The advent of  new Generation Wireless System (such as the 5G and, in perspective, 6G) is radically reshaping the business opportunities of mobile network operators. With the enhancement of several network capabilities, including bandwidth, latency, and data rate, the 5G technology allows the implementation of several use-cases not viable  with previous mobile communication technologies. Indeed, in addition to providing voice and data connectivity services, the 5G technology supports new applications such as vehicle to vehicle communication, industrial automation, smart cities, and health applications, just to cite a few. The introduction of these new applications has led to a new call for evaluating the security of the technologies, architectures, and scenarios involved in this new mobile communication landscape~\cite{8668635}. 
Although the major threats are mostly the same of current/previous communication technologies (jamming, DDoS, MITM, eavesdropping, etc.), the 5G environment demands the study of new requirements for the development of effective countermeasures. In such a scenario, puzzles could play a key role in providing powerful features, already consolidated in previous technologies, to the new security mechanisms developed for the 5G architecture.\\
The performance that 5G can guarantee enable the ultra-reliable and ultra-low latency services required by the Vehicular Ad-hoc Networks (VANETs) to implement road safety, intelligent traffic management, information dissemination among vehicles, automatic driving, etc. 
For the security of these services, however, the reliability and authenticity of the information exchanged between vehicles, as well as the ones exchanged between vehicles and road infrastructure, becomes crucial. For this reason, several authentication schemes have been proposed to protect communications within VANETs. Regardless of their implementation details, all the proposed protocols suffer from DoS attacks that, in the low latency and high bandwidth 5G architecture, can be easily performed. To mitigate this problem, a puzzle-based co-authentication scheme has been presented in~\cite{8337735}. The main goal of this proposal is to make the number of authentication requests that a vehicle can generate in a specific time interval, less than or equal to the number of requests that a legitimate vehicle can verify in the same amount of time. The authors modeled their puzzle as a variant of the Hashcash~\cite{back2002hashcash} scheme, carefully adjusting the difficulty level to achieve the aforementioned condition. 
Although promising, this solution has several limitations. First, the proposed countermeasure are effective under the assumption that the attacker has the same computational power of the legitimate vehicles. An attacker equipped with more powerful hardware could solve the puzzles efficiently, easily bypassing the countermeasure. \\
The use of a CPU-bound scheme in 5G networks could also lead to other problems. The energy consumption needed for the computation of the solution, for example, could be a serious disadvantage, especially for resource constraint devices typically used in mobile networks. Besides, bandwidth-bound puzzles may also be ineffective in this new network architecture due to the increase in bandwidth available for individual devices in the 5G environment. Indeed, further research efforts are needed to identify all these limitations and collect new specifications to design a new generation of puzzle schemes tailored to the peculiarities of the 5G network architecture.
Furthermore, while the deployment of 5G is at its beginning, 
the scientific community already began the design of the next generation of mobile networks: the 6G. This technology, although still in its design stage, is expected to merge the physical and the digital world by providing full support to a large variety of sensors, thus enabling many  new use cases~\cite{Kantola2019}. In this context, the security risks will grow as the possible threats cross from the digital to the physical world, magnifying the consequences of possible successful attacks. This scenario increases the need to develop effective defenses and, as already mentioned, puzzles could be very useful to this end.
\subsection{Quantum and Post-quantum Era}\label{sub:post-quantum}
Classical computers perform logical operations and store data relying on the definite position of individual bits, represented as binary states 0 and 1. Quantum computing, instead, makes use of quantum mechanical phenomena to manipulate and store data, acquiring the potential to process exponentially more data compared to classical computers. The potential danger posed to IT security by quantum computing was first established in 1994, by a US mathematician and computer scientist, Peter Shor. He published a quantum computer algorithm~\cite{shor1994algorithms} able to theoretically break, in a matter of seconds, some of the most used encryption techniques previously assumed secure.\\
The vast majority of puzzles are resistant to the threat introduced by this technology. In fact, quantum computing only increases the computational power of an attacker, with no effect on puzzles bounded to resources other than the CPU, such as the memory or the network. The only category of puzzle that could be exposed to this threat is, therefore, the CPU-bound schemes. However, not all CPU-bound puzzles will be compromised by quantum computing. Indeed, several works demonstrated that quantum computers are capable of solving complex problems unfeasible for classic computers only by using algorithms that exploit the power of quantum parallelism. For example, a quantum computer cannot be faster than a standard one in multiplications~\cite{Mavroeidis2018}. 
Quantum computers could be used to efficiently solve some problems underlying the asymmetric cryptography, such as the large prime integer factorization and the discrete logarithm problem, while they could be not so efficient in computing the pre-image of a hash function, or in generating a collision. Consequently, the hash-based puzzles can be safely used in the post-quantum era as long as they use a hash function that provides an output with an adequate length, such as sha-2 and sha-3~\cite{Mavroeidis2018}. Further research efforts are needed in this field to evaluate the security of the non-hash-based puzzles in the quantum world and, if necessary, make them resistant against quantum computation. 
The first effort in this direction has been made by Brassard et al. in~\cite{Brassard2011, 4455936}, where the authors provided a quantum-resistant key establishment scheme based on Merkle puzzle. The field though, from the research perspective, is still an open and exciting one.
\subsection{Location-Based Services}\label{sub:LBS}
The new generation of mobile networks, together with the mobile devices market, is also pushing the spreading of a new category of services, called Location-Based Services (LBSs). LBSs provide users with accurate and targeted information based on their geographic location, enabling a wide range of use cases, especially in VANETs. In this context, the validation of the real position of a client requesting access to an LBS service becomes of primary importance to ensure the security and stability of the entire system. Several mechanisms, called proof-of-location, have been proposed to solve this problem. The proof-of-location aims to certify the presence of a device in a specific geographical coordinate, at a particular instant of time. Puzzles could be used in this context to allow a prover to demonstrate his geographic location to a verifier. Researchers have started  studying  this scenario in~\cite{8431966, brambilla2016using}, proposing a blockchain-based proof-of-location mechanism to ensure location trustworthiness without compromising user privacy.}

\section{Conclusion}\label{conclusion}
In this paper, we have provided an extensive introduction to the concept of `\textit{puzzle}' and presented the evolution 
of this notion. We have studied the various types of puzzles from different perspectives and classified them 
based on the way they are applied, 
how the solution is used, and the required resources that bound their solving time.  
Based on our extensive review on the applications of puzzles, we have identified the key requirements that must be satisfied for a puzzle to be applied in a particular field. We have also investigated the influence and impact of puzzles, and carved out the elements that hinder 
their effectiveness in each of the addressed fields.
Moreover, we have provided a thorough review of the different types of construction schemes including CPU-bound, memory-bound, memory-hard, and bandwidth-bound schemes. For each type, we examined the different approaches and techniques used in designing the puzzle and highlighted their distinctive characteristics. We have also identified the limitations and benefits provided by each technique. 
Further, we have also provided
a few research directions that may lead to new insights in the field of puzzles.
Finally this survey, other than being interesting on its own, can be also used as a  guideline for 
designing effective puzzles for a wide range of applications. 

\section*{Acknowledgments}
The  authors  would  like  to  thank  the  anonymous  reviewers for their comments and suggestions, that helped improving the quality of the manuscript. \\*
This publication was partially supported by awards NPRP 11S-0109-180242,  UREP 23-065-1-014,  and  
NPRP  X-063-1-014 from the QNRF-Qatar National Research Fund, a member of The  Qatar  Foundation.  The  information  and  views  set  out  in this publication are those of the authors and do not necessarily reflect the official opinion of the QNRF.


\bibliographystyle{ACM-Reference-Format}
\bibliography{references.bib}

\end{document}